\documentclass[a4paper,11pt]{article}
\pdfoutput=1 
\usepackage{jheppub}  
\usepackage{graphicx,color}
\usepackage{amsmath}
\usepackage{autobreak}
\allowdisplaybreaks
\newcommand{\ben}{\begin{enumerate}}
\newcommand{\een}{\end{enumerate}}
\newcommand{\beq}{\begin{equation}}
\newcommand{\eeq}{\end{equation}}

\def\g0#1bbH{{g^{b(#1)}_{_{d,0}} }}
\usepackage[dvipsnames]{xcolor}

\newcommand{\NF}{n_f}

\newcommand{\NB}{\overbar{N}}
\def\zo{\overline{z}_1}
\def\zt{\overline{z}_2}
\newcommand{\calX}{{\cal X}}

\newcommand{\calO}{{\cal O}}
\newcommand{\calD}{{\cal D}}

\newcommand{\zb}{{\bar{z}}}
\newcommand{\Lomg}{L_\omega}
\newcommand{\omgot}{\omega_{12}}
\def\g{\overline {\cal G}}

\def\zo{\overline{z}_1}
\def\zt{\overline{z}_2}

\def\bt#1{{\beta_{#1}}}

\newcommand{\eq}[1]{eq.\ (\ref{#1})}
\newcommand{\fig}[1]{fig.\ (\ref{#1})}
\newcommand{\tab}[1]{tab.\ (\ref{#1})}
\newcommand{\sect}[1]{sec.\ (\ref{#1})}

\definecolor{amber}{rgb}{1.0, 0.49, 0.0}
\newcommand{\nn}{\nonumber\\}
\newcommand{\df}{{\rm{d}}}
\newcommand{\mur}{\mu_R}
\newcommand{\muf}{\mu_F}

\newcommand{\overbar}[1]{\,\overline{\!{#1}}}

\newcommand{\as}{a_s}
\newcommand{\ash}{\widehat{a}_s}
\newcommand{\als}{\alpha_s}
\newcommand{\lam}{\lambda}

\usepackage{bm}
\SetSymbolFont{largesymbols}{bold}{OMX}{txex}{b}{n}
\newcommand{\app}[1]{\ (\ref{#1})}

\usepackage{array}
\newcolumntype{P}[1]{>{\centering\arraybackslash}p{#1}}
\newcommand{\nsv}[1]{$\overline{\rm #1}$}
\usepackage[normalem]{ulem}

\title{\bf Next-to-soft threshold effects on Higgs boson production via bottom quark annihilation}
\author[a]{\bf Goutam Das}
\author[b,c]{\bf and Aparna Sankar} 
\affiliation[a]{
        Institut f{\"u}r Theoretische 
        Teilchenphysik und Kosmologie,\\
        RWTH Aachen University, 
        D-52056 Aachen, Germany
}
\affiliation[b]{
Physik Department T31, James-Franck-Stra{\ss}e 1,\\ Technische Universit{\"a}t M{\"u}nchen,
D-85748 Garching, Germany
}
\affiliation[c]{
Max-Planck-Institut f\"ur Physik, Boltzmannstraße 8, 85748 Garching, Germany
}

\emailAdd{goutam@physik.rwth-aachen.de}
\emailAdd{aparna@mpp.mpg.de}
\abstract{We examine the behavior of singular leading and sub-leading logarithms, commonly referred to as the soft-virtual (SV) and next-to-soft-virtual (NSV) terms, in the production of the Higgs boson via the bottom quark annihilation channel. We derive analytic expressions at the SV+NSV resummed level, up to next-to-next-to-next-to-leading logarithmic (N3LL) accuracy in perturbative QCD, applicable to both the inclusive production cross-section and rapidity distribution. This has been achieved using an existing resummation formalism based on factorization and renormalization group (RG) invariance. A phenomenological analysis of our resummed results is performed for the 13 TeV Large Hadron Collider. Furthermore, we present the SV coefficients at the fourth order in the strong coupling and explore the impact of next-to-next-to-next-to-next-to-leading logarithmic (N4LL) resummation on the total cross-section.} 
\begin{document} 
%%%Preprint
\preprint{
TTK-24-29, 
P3H-24-056\\
\vspace{1.0cm}
\hspace{10.9cm}MPP-2024-167
\vspace{-4cm}
}

\maketitle
%%%Introduction
\section{Introduction} \label{sec:INTRODUCTION}
The discovery of the Higgs boson at the Large Hadron Collider (LHC) has firmly established the Standard Model (SM) of particle physics. This landmark achievement has shifted the focus of the LHC's research program toward the precise measurement of Higgs boson properties. In the current and the forthcoming LHC runs, a primary objective will be to accurately measure the interactions of the Higgs boson with other fundamental particles. Precision calculations are critical in these efforts, providing higher-order contributions in perturbation theory and enhancing the accuracy of predictions related to the Higgs boson. The exploration of the Higgs sector is essential not only for validating the SM but also for identifying potential new physics phenomena. To date, the measured couplings of the Higgs boson to top and bottom quarks, $W^\pm$, and $Z$ bosons, and tau leptons have been consistent with the SM predictions. However, as the LHC continues to accumulate data, the precision of these measurements increases, thereby improving sensitivity to any deviations from the SM. Moreover, couplings that are currently constrained by large statistical uncertainties, such as the Higgs boson's self-interaction, are anticipated to become more accessible soon.

At the LHC, Higgs bosons are predominantly produced via the gluon fusion channel, making it the primary focus of theoretical studies \cite{Graudenz:1992pv,Spira:1995rr,Harlander:2002wh,Anastasiou:2002yz,Ravindran:2003um,Anastasiou:2008tj,Marzani:2008az,Pak:2009dg,Harlander:2009my,Harlander:2009mq,Bolzoni:2010xr,Bolzoni:2011cu,deFlorian:2012za,Anastasiou:2015vya,Dreyer:2016oyx,Mistlberger:2018etf,Das:2020adl,Buckley:2021gfw,Czakon:2021yub}. Nevertheless, the Higgs production via the bottom quark annihilation channel, though sub-dominant also deserves investigation \cite{Dicus:1998hs,Balazs:1998sb,Maltoni:2003pn,Harlander:2003ai,Dittmaier:2003ej,Dawson:2003kb,Belyaev:2005bs,Harlander:2010cz,Harlander:2011fx,Buhler:2012ytl,Harlander:2012pb,Harlander:2014hya,Wiesemann:2014ioa,Ahmed:2014cha,Bonvini:2015pxa,Bonvini:2016fgf,Forte:2015hba,Forte:2016sja,Harlander:2015xur,Lim:2016wjo,Das:2023rif,Gehrmann:2014vha,Chakraborty:2022yan,Duhr:2019kwi,Duhr:2020kzd,Mondini:2021nck,Cal:2023mib}. As precise measurements of the Higgs cross-sections are pursued at the LHC, incorporating bottom quark-initiated channels into theoretical predictions becomes essential. The Yukawa coupling of the bottom quark is particularly noteworthy, as several New Physics models, such as minimal super-symmetric extensions of the Standard Model, predict enhanced bottom-Yukawa couplings. The interactions between the Higgs boson and the bottom quark can be probed at the LHC either through the Higgs decaying into a pair of bottom quarks or through its production from bottom quarks. While it is theoretically possible to directly constrain the bottom quark Yukawa coupling by measuring the decay of a Higgs boson into a bottom quark pair, this measurement is challenging at a hadron collider due to the purely hadronic final state signature, despite the decay's large branching fraction \cite{CMS:2018nsn,ATLAS:2018kot}. Moreover, the accurate measurements of Higgs boson decays at the LHC also rely on precise predictions of its inclusive production cross-section through bottom quark annihilation. Therefore, studying Higgs production processes involving bottom quarks is advantageous. The objective of this paper is thus to conduct a phenomenological study of Higgs boson production through bottom quark fusion at the LHC.

Given the bottom quark's small but non-negligible mass, theoretical models for LHC processes involving bottom quarks typically employ two different approaches: the four-flavor scheme (4FS) and the five-flavor scheme (5FS). In the 5FS, the bottom quark is treated as a massless parton, with finite-mass effects neglected except for collinear logarithms resummed into the parton density functions and in the Yukawa coupling. This approach simplifies the computation of higher-order corrections in the strong coupling constant, as all quark species are considered massless (with top quark effects neglected). The inclusive Higgs production cross-section in bottom quark fusion has been computed in this scheme up to the next-to-next-to-next-to-leading order (N3LO) \cite{Dicus:1998hs,Balazs:1998sb,Maltoni:2003pn,Harlander:2003ai,Duhr:2019kwi}. Analyses of pure QED and mixed QCD-QED effects indicate that the resulting corrections are below 0.03\% of the leading order (LO) \cite{AH:2019xds}. On the other hand, the 4FS treats the bottom quark as massive, which leads to higher final-state multiplicities. Therefore, Higgs production in bottom quark fusion is currently known only through next-to-leading order (NLO) in this scheme \cite{Dittmaier:2003ej,Dawson:2003kb,Wiesemann:2014ioa,Deutschmann:2018avk}. Numerous studies have explored the differences between 4FS and 5FS results and consistent combination of predictions in two schemes has been obtained in various works \cite{Aivazis:1993pi,Cacciari:1998it,Forte:2010ta,Harlander:2011aa,Maltoni:2012pa,Bonvini:2015pxa,Forte:2015hba,Bonvini:2016fgf,Forte:2016sja,Lim:2016wjo,Duhr:2020kzd} through different matching procedures.

Beyond the fixed-order, there have been studies to incorporate the leading soft-virtual effects at third-order \cite{Ahmed:2014cha} and at next-to-next-to-next-to-leading logarithmic (N3LL) accuracy \cite{AH:2019phz,Das:2022zie}. Beyond leading power, there have been interests in the next-to-leading power threshold studies \cite{Moch:2009hr,Soar:2009yh,Bonocore:2015esa,Bonocore:2016awd,DelDuca:2017twk,Bahjat-Abbas:2019fqa,Beneke:2019oqx,vanBeekveld:2019cks,Das:2020adl,AH:2020iki,vanBeekveld:2021hhv,AH:2022lpp}. In \cite{AH:2020iki,AH:2022lpp} explicit analytic results for the next-to-soft threshold corrections as well as a general expression for the resummation have been presented to N3LO using the principles of renormalization group (RG) invariance and factorization. Recently, in \cite{Biello:2024dha,Biello:2024pad,Biello:2024vdh}, a matching of next-to-next-to-leading order (N2LO) predictions with parton shower (N2LO+PS) has been performed in the 5FS.

Besides the inclusive production cross-section, the differential rapidity distribution of the Higgs boson in the bottom annihilation channel has been studied to N2LO \cite{Buhler:2012ytl,Mondini:2021nck}. There have been efforts to improve this accuracy with threshold corrections to third order and beyond \cite{Ahmed:2014era,Ahmed:2020amh}. Further, a computation of next-to-soft virtual (NSV) corrections for rapidity has been provided in \cite{AH:2020qoa}. Recently, in \cite{Das:2023rif}, a phenomenological study of soft gluon resummation effects has been performed at N3LL for Higgs rapidity in bottom annihilation, and a flat correction of $-2.5\%$ has been observed over the whole rapidity region.

In this article, we study the effect of soft and next-to-soft gluon effects on the total cross-section and rapidity distribution of the Higgs boson production in bottom quark annihilation. For this, we use the approach of well-established SV threshold resummation \cite{Sterman:1986aj,Catani:1989ne,Catani:2003zt,Manohar:2003vb,Eynck:2003fn,Moch:2005ba,Moch:2005ky,Laenen:2005uz,Ravindran:2005vv,Ravindran:2006cg,Ravindran:2006bu,Ravindran:2007sv,Idilbi:2006dg,Becher:2006mr} and recently introduced NSV resummation in \cite{AH:2020iki,AH:2020qoa,AH:2022lpp} for resumming the large logarithms for the total cross-section and rapidity distribution, respectively. For rapidity resummation, we follow the double Mellin approach introduced in \cite{Catani:1989ne,Westmark:2017uig,Banerjee:2017cfc,Banerjee:2018vvb,Banerjee:2018mkm,Ahmed:2020amh}.\footnote{There exist other approaches for rapidity resummation, eg.\ \cite{Bonvini:2010tp,Becher:2007ty,Lustermans:2019cau} and see \cite{Bonvini:2023mfj} for comparison among them.}

The article is organized as follows: in section (\ref{sec:THEORY}), we outline the theoretical framework for resummation of SV and NSV logarithms in the context of both the total cross-section and the rapidity distribution of Higgs boson production via bottom quark annihilation. Additionally, we present partial N4LO soft and next-to-soft results for the total cross-section. In section (\ref{sec:NUMERICS}), we present phenomenological analyses of SV and NSV resummation for the $13$ TeV LHC. Finally, we conclude our findings in \sect{sec:CONCLUSION} and collect relevant formulas in the appendix. 

%%%Theoretical framework%%%%%%%%%%%%%%%%%%%%%%%%%%%%
\section{Theoretical Framework} \label{sec:THEORY}
%%%%%%%%%%%%%%%%%%%%%%%%%%%%%%%%%%%%%%%%%%%%%%%%%%%%
The effective Lagrangian describing the interaction between a scalar Higgs boson ($\phi(x)$) and a bottom quark ($\psi_{b}(x)$) is expressed as
\begin{align}\label{eq:LAGRANGIAN}
{\cal L}_{\rm int}^{\rm (S)} 
= -\lambda_{y}~ \overbar{\psi}_{b}(x)\psi_{b}(x)\phi(x) \,.
\end{align}
The parameter $\lambda_y$ denotes the Yukawa interaction, which is defined as $\lambda_y = m_b/v$, where $v$ is the vacuum expectation value (VEV), and $m_b$ is the mass of the bottom quark. We adopt the five-flavor scheme (5FS), wherein the bottom quark is treated as massless except in the Yukawa coupling. In QCD improved parton model, the inclusive production of  Higgs boson through bottom quark annihilation at hadron collider can be expressed as
\begin{align}\label{eq:XSECT-MASTER}
\sigma(\tau) = 
\sum_{ i,j = \{q,\overline q,g\}} 
&\int_0^1 \df x_1 \int_0^1 \df x_2~
f_i(x_1,\mu_F) f_j(x_2,\mu_F) 
\int_0^1 \df z ~ \delta(\tau - x_1 x_2 z  )~
\widehat{\sigma}_{{\calX},ij}\left({z},\mu_F\right) \,,
\end{align}
where $\tau=m_H^2/S$ is the hadronic scaling variable with $m_H$ being the Higgs mass and $\sqrt{S}\equiv E_{\rm CM}$ being the hadronic centre-of-mass energy. $f_i(x_1,\muf)$ and $f_j(x_2,\muf)$ are the parton distribution functions (PDFs) of incoming partons $i$ and $j$ with momentum fractions $x_1$ and $x_2$, respectively and are renormalized at the factorization scale $\mu_F$. $\widehat{\sigma}_{{\calX},ij}$ are the partonic coefficient functions obtained from the partonic subprocesses after mass factorization at the scale $\mu_F$ and are calculable order by order in QCD perturbation theory in powers of $a_s=\alpha_s/(4 \pi)$ with $\als$ being the strong coupling constant. Similarly, the differential distribution in the Higgs rapidity can be expressed analogous to \eq{eq:XSECT-MASTER} as,
\begin{align}\label{eq:RAPIDITY-MASTER}
        {\df\sigma(\tau, y)\over \df y } 
        =
        % \sigma_{0}(x_1^0,x_2^0,m_b,\mur) 
        \sum_{i,j= \{q,\bar{q},g\}}
        &\int_{0}^1 {\df  x_1}
        \int_{0}^1 {\df  x_2}~
        f_i\left({x_1},\muf\right)  
        f_j\left({x_2},\muf\right)
\nn  \times &
        \int_{0}^1 {\df  z_1}
        \int_{0}^1 {\df z_2}~
        \delta(x_1^0 - x_1 z_1)
        \delta(x_2^0 - x_2 z_2)~
\widehat{\sigma}_{{d},ij} (z_1,z_2,\muf)\,,
\end{align} 
where $\widehat{\sigma}_{{d},ij}$ are the rapidity-dependent partonic coefficient functions. The scaling variables $x_{1(2)}^0$ are related to $\tau$ by $\tau= x_1^0 x_2^0 $ and to the hadronic rapidity $y$ of the Higgs boson by  $y=\frac{1}{2}\ln \left(\frac{x_1^0}{x_2^0}\right)$. 

The partonic coefficient functions appearing in \eq{eq:XSECT-MASTER} and \eq{eq:RAPIDITY-MASTER} in the case of Higgs production in bottom annihilation can be decomposed as follows (suppressing the $z(z_1,z_2)$-dependencies),
\begin{align}\label{eq:partonic-decompose}
\widehat{\sigma}_{{\cal O},ij}(\muf)
=
\sigma_{b \bar{b}}^{(0)}(\mur) 
\Big( 
	\delta_{ib} \delta_{j\bar{b}}\Delta_{{\calO},b\bar{b}}^{\rm SV}\left(\muf,\mur\right) 
      + \Delta_{{\calO},ij}^{\rm REG}\left(\muf,\mur\right)
\Big) ~ \text{ with }  {\calO} = \{ \calX, d \} \,,
\end{align}
where $\sigma_{b \bar{b}}^{(0)}(\mur)$ is the overall Born normalization factor and is given as,
\begin{align}
\sigma_{b \bar{b}}^{(0)}(\mur) 
=
\frac{\pi \lam_y^{2}(\mur) \tau}{6 m_H^2}
\equiv
\frac{\pi m_b^{2}(\mur) \tau}{6 m_H^2 v^2} \,.
\end{align}
In \eq{eq:partonic-decompose}, $\Delta^{\rm SV}_{\calO,b\bar{b}}$ (known as the soft-virtual (SV) coefficient function) is singular in the threshold limit whereas $\Delta_{{\calO},ij}^{\rm REG}$ (regular coefficient function) is finite in the same limit. The SV terms get contributions from soft radiation and from the virtual form factor. Their structure has been studied widely in the literature 
\cite{Sterman:1986aj,Catani:1989ne,Catani:2003zt,Manohar:2003vb,Eynck:2003fn,Moch:2005ba,Moch:2005ky,Laenen:2005uz,Ravindran:2005vv,Ravindran:2006cg,Ravindran:2006bu,Ravindran:2007sv,Idilbi:2006dg,Becher:2006mr}. The regular contributions (REG), while being formally subdominant in the threshold region, can be important away from the threshold. We further decompose the SV and REG terms for the inclusive cross section as,
\begin{align}\label{eq:SV-REG-EXPANSION-XSECT}
\Delta^{\rm SV}_{\calX,b\bar{b}}(z) 
&= \sum_{n=0}^{\infty}\as^n
\left( 
\Delta^{(n)}_{\calX, \delta} \delta(\zb) + 
\sum_{k=0}^{2n-1}  \Delta^{(n)}_{\calX, {\cal D}_k} {\cal D}_k(\zb) 
\right)\,,
\nn
\Delta^{\rm REG}_{\calX,ij}(z) 
&= \sum_{n=0}^{\infty}\as^n
\left( 
\sum_{k=0}^{2n-1} \Delta_{\calX,ij,{\ln}_k}^{(n)} \ln^k(\zb) 
+
\Delta_{\calX,ij}^{(n)}(z)
\right)\,,
\end{align}
where $\zb = 1-z$ and ${\cal D}_k(\zb)=\big[{\ln^i\zb \over \zb}\big]_+$  are the plus-distributions. Note that $\Delta_{\calX,ij,{\ln}_k}^{(n)}$ are the coefficients of the next-to-soft (NSV) terms ($\ln^k (\zb)$) which originate from the soft gluon radiations. These NSV terms are subdominant compared to the SV ones. The term $\Delta_{\calX,ij}^{(n)}$ in \eq{eq:SV-REG-EXPANSION-XSECT} originates from the hard radiations and vanishes in the threshold limit $z \to 1$.

Similarly, for the rapidity distribution, we decompose SV and REG contributions as 
\begin{align}\label{eq:SV-REG-EXPANSION-RAP}
\Delta^{\rm SV}_{{d},b\bar{b}} (z_1,z_2) =&\sum_{n=0}^\infty a_s^n 
\Bigg(\Delta^{(n)}_{d, \delta \overline{\delta}} \delta(\zo) \delta(\zt)
+\sum_{j=0}^{2n-1} \Big(
 \Delta^{(n)}_{d, \delta \overline{{\cal D}}_j}\delta(\zo) \overline{{\cal D}}_j
+\Delta^{(n)}_{d, \overline{\delta} {\cal D}_j} \delta(\zt) {\cal D}_j
\Big)
\nn &
+ \sum_{j=0}^{2n-2} \sum_{k=0}^{2n-2-j} \Delta^{(n)}_{d, {\cal D}_j \overline{{\cal D}}_k}
 {\cal D}_j  \overline{{\cal D}}_k
 \Bigg)\,,
\nn
\Delta^{\rm REG}_{d,ij}(z_1,z_2) 
=& \sum_{n=0}^{\infty}\as^n
\Bigg( 
\sum_{k=0}^{2n-1} \Big(
 \Delta_{d,ij, \delta \overline{L}_k}^{(n)} ~ \delta(\zo) \ln^k(\zt) 
+\Delta_{d,ij, \overline{\delta} L_k}^{(n)}~ \delta(\zt) \ln^k(\zo) 
\Big)
\nn & 
+ \sum_{j=0}^{2n-2} \sum_{k=0}^{2n-2-j} \Big(
   \Delta_{d,ij, {\cal D}_j \overline{L}_k}^{(n)}~ {\cal D}_j \ln^k(\zt) 
+  \Delta_{d,ij, \overline{{\cal D}}_j L_k}^{(n)} ~\overline{{\cal D}}_j \ln^k(\zo)   
\Big)
+ \Delta_{d,ij}^{(n)}(z_1,z_2)
\Bigg)\,,
\end{align}
with $\overline{z}_l= 1-z_l\,, \overline{\delta} = \delta(\zt)\,, {\cal D}_i=\Big[{\ln^i \zo \over \zo}\Big]_+\,, \overline{{\cal D}}_i=\Big[{\ln^i\zt \over \zt}\Big]_+\,, L_k = \ln^k(\zo) \,, \overline{L}_k = \ln^k(\zt)\,.$ The structure of the NSV component in the rapidity distribution is more involved compared to the inclusive cross-section case due to the presence of terms like $\delta( \overline{z}_l) \ln^k(\overline{z}_m)$, ${\cal D}_j \ln^k(\overline{z}_l)$ and $\overline{{\cal D}}_j \ln^k(\overline{z}_m)$. As before, the last term $\Delta_{d,ij}^{(n)}(z_1,z_2)$ vanishes in the threshold limits ($\overline{z}_l \to 0$). We also remind that while SV corrections originate purely from the diagonal $b \bar{b}$ channel, the NSV terms additionally receive contributions from the off-diagonal channel. However, we only study the contributions originating from the diagonal channel, for which the resummation formalism is now well-established \cite{AH:2020iki}. For convenience, we combine the SV and NSV terms in the diagonal channel and denote it by $\Delta^b_{\calO}$ (suppressing the $z(z_1,z_2)$-dependencies), 
\begin{align}\label{eq:SV-NSV}
\Delta^b_{\calO}(\mu_R^2,\mu_F^2) = \Delta_{\calO, b \bar{b}}^{\rm SV}(\mu_R^2,\mu_F^2) +
\Delta_{\calO, b \bar{b}}^{\rm NSV}(\mu_R^2,\mu_F^2)\,,~~ \text{ with } \calO \in \{ \calX, d \} \,.
\end{align}
We briefly review the computation of $\Delta^b_{\calO}$ for the cases of both inclusive cross-section and rapidity distribution in the following subsections.
%%%%%%%%%%%%%%%%%%%%%%%%%%%%%%%%%%%%%%%%%%%%%%%%%%%%%%%%
\subsection{Inclusive production}\label{sec:incl-xsect}
%%%%%%%%%%%%%%%%%%%%%%%%%%%%%%%%%%%%%%%%%%%%%%%%%%%%%%%%
We begin with discussing $\Delta^b_{\calX}$ for the inclusive production of the Higgs boson in bottom annihilation. In dimensional regularisation ($d=4-2\varepsilon$), the total SV+NSV contribution for the inclusive cross section can be formulated \cite{AH:2020iki} in terms of a finite function $\Psi^b_{\calX}$ as follows,
\begin{align}\label{eq:MASTER-XSECT-EXP}
\Delta^b_{\calX}(\mu_R^2,\mu_F^2,z) = 
\mathcal{C}\exp \bigg( \Psi^b_{\calX}\big(\mu_R^2,\mu_F^2,z,\varepsilon\big)\bigg)\bigg |_{\varepsilon=0} \,.
\end{align}
The symbol $\cal C$ implies a Mellin convolution ($\otimes$) with the following expansion,
\begin{align}
    {\cal C} e^{f(z)} = \delta(\zb) + \frac{1}{1!}f(z) + \frac{1}{2!} f(z) \otimes f(z) + \cdots
\end{align}
In the threshold limit, $\Psi^b_{\calX}$ follows \cite{AH:2020iki} a factorization formula given by,
\begin{align}\label{eq:MASTER-XSECT-FAC}
    \Psi^b_{\calX}\big(\mu_R^2,\mu_F^2,z,\varepsilon\big) =&
    \Bigg( 
        \ln \big( Z_{b}\big(\ash,\mu^2,\mur^2,\varepsilon\big)\big)^2 +   \ln \big| \widehat{F}_{b}\big(\ash,\mu^2,-m^2_H,\varepsilon\big)\big|^2
        \Bigg) \delta(\zb)
    \nn
    &+2 \mathrm{\Phi}_b\big(\ash,\mu^2,m^2_H,z,\varepsilon\big) 
    - 2{\cal C} \ln \Gamma_{b \bar{b}}\big(\ash,\mu^2,\muf^2,z,\varepsilon\big) \,.
\end{align}
Here $Z_b$ is the UV renormalization constant \cite{vanRitbergen:1997va,Czakon:2004bu,Baikov:2014qja,Baikov:2017ujl} for the bottom Yukawa coupling $\lambda$, $\widehat{F}_{b}$ is the bare Higgs form factor in bottom annihilation \cite{Gehrmann:2014vha,Chakraborty:2022yan}, $\mathrm{\Phi}_b$ is the soft distribution function \cite{AH:2020iki} originating from the soft gluon radiation and $\Gamma_{b \bar{b}}$ is the mass factorization kernel \cite{Ellis:1996nn,Soar:2009yh,Vogt:2004mw,Moch:2004pa,Ablinger:2017tan,Moch:2017uml,Dixon:2019lnw,Chen:2020uvt,Falcioni:2023luc,Falcioni:2023vqq,Falcioni:2024xyt,Guan:2024hlf} which removes the initial state collinear singularities. Each of these functions is divergent and these divergences are regulated in the dimensional regularisation and their sum is finite. The scale $\mu$ is introduced to make the bare coupling $\ash$ dimensionless in $d$-dimension. 

In \cite{AH:2020iki,AH:2022lpp}, an integral representation for the function $\Psi^b_{\calX}$ was obtained which takes the following form,
\begin{align}
\label{eq:resumz}
\Delta^b_{\calX}(\mu_R^2,\mu_F^2,z)= C^b_0(\mu_R^2,\mu_F^2)
~~{\cal C} \exp \Bigg(2 \Psi^b_{\calX, \cal D} (\mu_F^2,z) \Bigg)\,,
\end{align}
with
\begin{align}
\label{eq:phicint}
\Psi^b_{\calX, \cal D} (\mu_F^2,z) &= {1 \over 2}
\int_{\mu_F^2}^{q^2 \zb^2} {\df \lambda^2 \over \lambda^2}
        P_{b \bar b} \big(a_s(\lambda^2),z\big)  
+\left[{1 \over \zb} \overline G^b_{SV}\big(a_s(q^2 \zb^2)\big)\right]_+ 
+ \varphi_{f,b}\big(a_s(q^2\zb^2),z\big).
\end{align}
The exponent ($\Psi^b_{\calX, \cal D}$) captures soft and next-to-soft terms in the threshold limit ($z\to 1$) whereas the prefactor $C_0^b$ is constant in the same limit. The function $\overline{G}^b_{SV}\big(a_s(q^2\zb^2),\epsilon\big)$ in \eq{eq:phicint} contributes to the SV resummation and contains the SV soft anomalous dimension ($f^b$) (see eg. \cite{Ravindran:2005vv,Ravindran:2006cg}). Naturally, it is related to the well-known threshold exponent $\textbf{D}^b\big(a_s(q^2\zb^2)\big)$ as shown in \cite{Ravindran:2006cg}. These are now known to the third order in QCD (see Appendix C of \cite{AH:2019phz}). The quark flavor-universality of these coefficients allows us to relate the fourth-order results to the same as that of the Drell-Yan process which are provided in \cite{Das:2020adl}. The splitting function $P_{b \bar b}$ appearing in \eq{eq:phicint} is the Altarelli-Parisi (AP) splitting function at the threshold and it takes the form,
\begin{align}
\label{eq:Pbb}
        P_{b \bar b}\big(z,a_s(\mu_F^2)\big) &= 
        2 \Bigg(
             A^b(a_s(\mu_F^2)) \calD_0(\zb)
            + B^b(a_s(\mu_F^2)) \delta(\zb)
            + C^b(a_s(\mu_F^2)) \ln(\zb) 
            + D^b(a_s(\mu_F^2)) \Bigg)\,.
\end{align}
The coefficient of the distribution $\calD_0(\zb)$ is the well-known cusp anomalous dimension ($A^b$) whereas the coefficient of the $\delta(\zb)$ distribution is the collinear anomalous dimension ($B^b$). Both of them contribute to the SV resummation. The coefficients $C^b$ and $D^b$ appear only in the NSV resummation. All these anomalous dimensions can be expanded in powers of $a_s(\mu_F^2)$ as,
\begin{align}
\label{eq:expand-cts}
X(a_s(\mu_F^2)) = \sum_{i=1}^\infty a_s^i(\mu_F^2) X_i, \quad \quad \textcolor{black}{X = A^b,B^b,C^b,D^b}\,,
\end{align}
All of these coefficients are available to the third order in QCD \cite{Moch:2004pa,Vogt:2004mw} for quite some time. The cusp anomalous dimensions are also available up to four loops \cite{Korchemsky:1987wg,Moch:2004pa,Vogt:2004mw,Henn:2019swt,vonManteuffel:2020vjv} and beyond \cite{Herzog:2018kwj} in QCD. The four-loop results of the soft ($f_4^b$ ) and collinear ($B_4^b$) anomalous dimensions can be obtained from \cite{Davies:2016jie,Das:2019btv,Das:2020adl,vonManteuffel:2020vjv}. The fourth-order NSV coefficients $C_4^b$ and $D_4^b$ can be found in \cite{Moch:2017uml}. The NSV function $\varphi_{f,b}$ appearing in \eq{eq:phicint} is given by
\begin{align}
\label{eq:varphiexp}
\varphi_{f,b}(a_s(q^2\zb^2),z) = \sum_{i=1}^\infty a_s^i(q^2\zb^2) \sum_{k=0}^i \varphi_{b,i}^{(k)} \ln^k(\zb) \,,
\end{align} 
where $q^2 = m^2_H$ and the coefficients $\varphi_{b,i}^{(k)}$ are known to third-order \cite{AH:2020iki,AH:2022lpp}. The NSV coefficients $\varphi_{c,i}^{(k)}$ for Higgs production in bottom annihilation ($c = b$) and the Drell-Yan process ($c = q$) are identical up to $i = 2$. Additionally, the diagonal coefficients $\varphi_{c,i}^{(i)}$ are also found to be the same for both processes which have been confirmed up to the third order in \cite{AH:2022lpp,AH:2020iki}.

The resummation of threshold logarithms is effectively performed in the Mellin space, where the limit $z \rightarrow 1$ corresponds to large $N$. The plus-distributions translate into large logarithms ($\ln N$) at the threshold spoiling the predictivity of a truncated fixed-order result. This large threshold effect can be taken into account by reorganizing the series using the integral representation \eqref{eq:phicint} effectively resumming these large logarithms to all orders. While in the SV resummation, one resums large logarithms of $\ln^k N$, in the case of the NSV resummation additionally subleading logarithms of the type $\ln^k N /N$ are resummed. The Mellin moment of $\Delta^b_{\calX}$ was obtained in \cite{AH:2020iki,AH:2022lpp} and is given by 
\begin{align}
\label{eq:DeltaN}
\Delta^b_{\calX,\NB }(\mu_R^2,\mu_F^2) = C_0^b(\mu_R^2,\mu_F^2) \exp\left(
\Psi_{\calX, \NB}^b (\mu_F^2,\textcolor{black}{\mu_R^2})
\right)\,,
\end{align}
where
\begin{align}\label{eq:Mellin}
\Psi_{\calX, \NB }^b(\mu_F^2,\textcolor{black}{\mu_R^2}) = 2 \int_0^1 dz z^{N-1}\Psi_{\calX, \cal D}^b (\mu_F^2,\textcolor{black}{\mu_R^2},z) .
\end{align}
Here,  we compute the Mellin moment in the large $\NB \equiv N\exp(\gamma_{E})$  limit meaning the entire $\gamma_{E}^{}$ dependent terms are also exponentiated through $\NB$. It has been observed in \cite{Das:2019btv} that the exponentiation of these spurious $\gamma_E$ makes the convergences faster. The exponent $\Psi_{\calX, \NB }^b$ contains both SV  and NSV terms and can be separated as follows,
\begin{align}
\label{eq:Psi}
\Psi_{\calX, \NB }^b = \Psi_{\calX, {\rm sv}, {\NB }}^b  + \Psi_{\calX, {\rm nsv}, \NB }^{b}\,.
\end{align}
Naturally, $\Psi_{\calX, {\rm{sv}},\NB }^b$ contains  terms of the form $\ln^j \NB, ~~j=0,1,\cdots$ while $\Psi_{\calX, {{\rm nsv}},\NB }^b$ contains terms like $ (\ln^j \NB) / N, ~~j=0,1,\cdots$.

The structure of SV resummed exponent $\Psi_{\calX, {\rm sv}, \NB}^b$ has been studied widely in the literature and it takes the following expansion,
\begin{align}
\label{eq:PsiSVN}
        \Psi_{\calX, {\rm{sv}},\NB }^b = \ln g_0^M(a_s(\mu_R^2)) + g_1^b(\lambda)\ln \NB + \sum_{i=0}^\infty a_s^i(\mu_R^2) g_{i+2}^b(\lambda) \,,
\end{align}
where $\lam = 2 \bt0 a_s(\mu_R^2) \ln \NB $ with $\beta_0=11/3 ~C_A-2/3~ n_f$. The constant $\ln g_0^M(a_s(\mu_R^2))$ arises due to the Mellin transformation of the plus-distributions and can be combined with the $N$-independent function $C^b_0$ in \eq{eq:DeltaN}. This results in an overall $N$-independent constant $g_0^b$ given by, 
\begin{align}
\label{eq:gtilde}
g_0^b(\mu_R^2, \mu_F^2) 
= C_0^b(\mu_R^2,\mu_F^2) \  g_0^M(a_s(\mu_R^2))
= \sum_{i=0}^\infty a_s^i(\mu_R^2) g^b_{0,i}\,.
\end{align}
Therefore the coefficients $g^b_{0,i}$ are constants containing process-dependent information through the form factor and universal constants arising from the Mellin transformation of soft terms. The leading-log (LL) SV resummation is obtained by keeping only the first term $g_1^b(\lambda)\ln \NB$ in the exponent and from the prefactor in \eq{eq:gtilde} whereas for subsequent higher order SV resummation, one needs to keep more terms in the strong coupling expansion from the exponent and the $N$-dependent prefactor from \eq{eq:gtilde}. The first four coefficients ($g_1^b \dots g_4^b$) needed for N3LL resummation can be found in \cite{AH:2019phz}. The coefficient $g_5^b$ needed for N4LL resummation can be found from the same for DIS \cite{Moch:2005ba,Das:2019btv,AH:2020xll} with the relation  $g_5^b(\lam) = g_5^{\rm DIS}(2\lam)$. It is worth noting that there is still an ambiguity on which parts are exponentiated and which parts are not, leading to different schemes for resummation. In the above formulation, $\ln \NB$ terms are exponentiated (known as $\NB$-scheme) whereas one can also rewrite the \eq{eq:PsiSVN} such that only $\ln N$ terms are exponentiated and the spurious $\gamma_E$ terms are combined with the constant prefactor in \eq{eq:gtilde} (known as the $N$-scheme). A detailed study at the SV resummed level has been performed to estimate these effects in \cite{Das:2019btv,AH:2020cok}. The exponent  $\Psi_{\calX, {\rm{nsv}}, \NB}^b$ for the NSV resummation has been discussed recently in \cite{AH:2020iki,AH:2020xll,AH:2021vdc,AH:2021kvg,AH:2022lpp} and it takes the following form,
\begin{align}
\label{PsiNSVN}
 \Psi_{\calX, {\rm{nsv}},\NB}^b = {1 \over N} 
\sum_{i=0}^\infty a_s^i(\mu_R^2) \bigg ( \bar g_{i+1}^b(\lambda)
+ h^b_{i}(\lambda,\NB) \bigg)\,.
\end{align}
Similar to the SV resummation, each term in the expansion of the strong coupling in the above summation determines the NSV resummation order which we denote for $n$-th order as $\overline{\rm{N{\it n}LL}}$. To emphasize, while for SV resummation at LL, one needs the $N$-dependent exponent ($g_1^b(\lambda)\ln \NB $)  in \eq{eq:PsiSVN} and prefactor ($g^b_{0,0}$) in \eq{eq:gtilde}, for NSV resummation at ${\rm \overline{LL}}$, one additionally needs NSV exponents $\bar g_{1}^b(\lambda)$ and $h^b_{0}(\lambda,\NB)$ in \eq{PsiNSVN}. Similarly for higher order resummation one needs to include successive terms for these coefficients. The functions $h^b_{i}(\lambda,\NB)$ have further expansion as,
\begin{align}\label{hNSV}
h^b_i(\lambda,\overbar{N}) = \sum_{k=0}^{i} h^b_{ik}(\lambda)~ \ln^k \overbar{N} \,.
\end{align}
The exponents $\bar g^b_i(\lambda)$ are determined entirely by the cusp anomalous dimension ($A^b$) and the function $\overline{G}^b_{SV}$ given in \eqref{eq:phicint}, making them universal. In contrast, the coefficients $h_{ik}^b(\lambda)$ depend on the NSV coefficients $\varphi_{b,i}^{(k)}$ in \eqref{eq:varphiexp}, as well as the $C^b$ and $D^b$ coefficients from $P_{b \bar b}$ in \eqref{eq:Pbb}. Due to their dependence on $\varphi_{b,i}^{(k)}$, the exponents $h_{ik}^b(\lambda)$ are not universal. We present these coefficients up to the third order in Appendix\app{App:NSV-INCLUSIVE-COEFF}.

Finally, the resummed results obtained in \eq{eq:DeltaN} can be matched to the fixed order results to account for the missing regular terms (beyond NSV). This matching can be achieved through the following matching formula,
\begin{align}\label{eq:matchincl}
\sigma^{\rm {N{\it n}LO+\overline {\rm N{\it n}LL}}}(\tau) &= 
\sigma(\tau)\big|_{\rm {N{\it n}LO}} +
\sigma_{b \bar{b}}^{(0)}(\mu_R)
\sum_{ij\in\{b,\bar{b}\}}
  \int_{c-i\infty}^{c+i\infty} \frac{\df N}{2\pi i} \tau^{-N} 
  f_{i,N}(\mu_F^2) f_{j,N}(\mu_F^2) 
  \nn
&\times \bigg( \Delta_{\calX,\NB}^b \big|_{\overline {\rm {N{\it n}LL}}} - {\Delta_{\calX,\NB}^b}\big|_{\rm {tr{\text -}N{\it n}LO}}     \bigg)  \,.
\end{align}
The matching procedure removes any double counting from the fixed order $\sigma(\tau)\big|_{\rm {N{\it n}LO}}$. The symbol $``{\rm {tr{\text -}N{\it n}LO}}"$ in the last term in the bracket describes that $\Delta_{\calX,\NB}^b$ has been truncated to the fixed order. In the above equation, $ \overline {{\rm N{\it n}LL}}$ stands for the resummation result at SV+NSV accuracy. To distinguish between SV and SV+NSV resummation, all along the paper, we denote the former by N${\it n}$LL and the latter by $\overline{\rm  N{\it n}LL}$ for the $n^{\rm th}$ level logarithmic accuracy. Therefore the same matching procedure should be realized for the SV resummation with $\overline{\rm N{\it n}LL}$ replaced by ${\rm N{\it n}LL}$. The Mellin space PDF ($f_{i,N}$) can be evolved using {\sc Qcd-Pegasus} \cite{Vogt:2004ns}. Alternatively, one can use the technique described in \cite{Catani:1989ne,Catani:2003zt} to directly deal with PDFs in the $z$ space at the cost of taking a derivative of PDFs. The contour $c$ in the Mellin inversion can be chosen according to {\it Minimal prescription} \cite{Catani:1996yz} procedure to avoid the Landau pole singularity. In section \ref{sec:NUMERICS-INCLUSIVE}, we present the phenomenological results for the inclusive Higgs production process in bottom annihilation at the LHC. We mainly focus on the impact of the resummed results at $\overline{\rm N3LL}$ on the N3LO inclusive cross-section. In addition, we also explore the numerical effect of N4LL resummation on the approximate N4LO results at SV accuracy. 

\noindent
{\bf Approximate N4LO coefficients:}

\noindent
For the complete N4LO SV resummation one needs to know all the distributions ($\calD_i$) along with $\delta(\zb)$ at the fourth order. In particular the coefficient of the lowest plus-distribution $\calD_0$ takes the following form \cite{Ahmed:2020nci}, 
\begin{align}
\begin{autobreak}
          
   \Delta_{\calX, b \bar b}^{{(4)}, \rm SV}\Big|_{{{\cal D}}_0} 
  
  =    n_f \frac{d_F^{abcd} d_F^{abcd}}{N_F}   \bigg\{
           768
          + \frac{43520}{9} \zeta_5
          + \frac{10624}{9} \zeta_3
          - \frac{2432}{3} \zeta_3^2
          - \frac{9088}{3} \zeta_2
          - 256 \zeta_2 \zeta_3
          + \frac{640}{3} \zeta_2^2
          - \frac{18944}{315} \zeta_2^3

       +  4 b^q_{4,d_F^{abcd}d_F^{abcd}}  
           
          \bigg\}

       - \frac{d_F^{abcd} d_A^{abcd}}{N_F}     \bigg\{ 2 f^q_{4, d_F^{abcd} d_A^{abcd}}        \bigg\}

       + C_F n_f^3   \bigg\{
           \frac{10432}{2187}
          - \frac{3680}{81} \zeta_3
          - \frac{3200}{81} \zeta_2
          + \frac{224}{45} \zeta_2^2
          \bigg\}

       + C_F C_A n_f^2   \bigg\{
          - \frac{898033}{2916}
          + \frac{608}{3} \zeta_5
          + \frac{87280}{81} \zeta_3
          + \frac{293528}{243} \zeta_2
          - \frac{608}{9} \zeta_2 \zeta_3
          - \frac{3488}{15} \zeta_2^2
          \bigg\}

       + C_F C_A^2 n_f   \bigg\{
           \frac{10761379}{2916}
          - \frac{29552}{27} \zeta_5
          - \frac{948884}{81} \zeta_3
          - \frac{9736}{9} \zeta_3^2
          - \frac{2418814}{243} \zeta_2
          + \frac{28064}{9} \zeta_2 \zeta_3
          + \frac{85312}{27} \zeta_2^2
          - \frac{52688}{189} \zeta_2^3
          \bigg\}

       + C_F C_A^2 n_f  \bigg\{
          -  b^q_{4,n_f C_F^3}  
          \bigg\}

       + C_F C_A^2 n_f   \bigg\{
          - 2 b^q_{4,n_f C_F^2 C_A}  
          \bigg\}

       + C_F C_A^2 n_f\bigg\{
          - \frac{1}{12}  b^q_{4,d_F^{abcd}d_F^{abcd}}   
          \bigg\}

       + C_F C_A^3    \bigg\{
          - \frac{28325071}{2187}
          + 3400 \zeta_7
          - \frac{49840}{9} \zeta_5
          + \frac{867584}{27} \zeta_3
          - \frac{4664}{3} \zeta_3^2
          + \frac{5761670}{243} \zeta_2
          + 832 \zeta_2 \zeta_5
          - \frac{119624}{9} \zeta_2 \zeta_3
          - \frac{301264}{45} \zeta_2^2
          + 576 \zeta_2^2 \zeta_3
          + \frac{334312}{315} \zeta_2^3
          \bigg\}

       + C_F C_A^3    \bigg\{
           \frac{1}{12} f^q_{4, d_F^{abcd} d_A^{abcd}} 
          \bigg\}

       + C_F^2 n_f^2    \bigg\{
          - \frac{309953}{729}
          + \frac{33056}{9} \zeta_5
          + \frac{124976}{81} \zeta_3
          + \frac{314240}{729} \zeta_2
          - \frac{79360}{27} \zeta_2 \zeta_3
          + \frac{6976}{27} \zeta_2^2
          \bigg\}

       + C_F^2 C_A n_f    \bigg\{
           \frac{5590667}{1458}
          - 37616 \zeta_5
          - \frac{1606660}{81} \zeta_3
          + \frac{11728}{3} \zeta_3^2
          - \frac{2713546}{729} \zeta_2
          + \frac{313888}{9} \zeta_2 \zeta_3
          - \frac{1920992}{405} \zeta_2^2
          - \frac{1312}{105} \zeta_2^3
          \bigg\}

       + C_F^2 C_A n_f  \bigg\{
           4 b^q_{4, n_f C_F^2 C_A}   
          \bigg\}

       + C_F^2 C_A^2    \bigg\{
           \frac{1571464}{729}
          + \frac{1005056}{9} \zeta_5
          + \frac{5327504}{81} \zeta_3
          - \frac{88640}{3} \zeta_3^2
          + \frac{6077552}{729} \zeta_2
          + 3072 \zeta_2 \zeta_5
          - \frac{3124160}{27} \zeta_2 \zeta_3
          + \frac{3124352}{405} \zeta_2^2
          + \frac{121888}{15} \zeta_2^2 \zeta_3
          - \frac{34496}{15} \zeta_2^3
          \bigg\}

       + C_F^3 n_f   \bigg\{
          - \frac{48157}{54}
          - \frac{130624}{3} \zeta_5
          - \frac{19520}{9} \zeta_3
          + \frac{106336}{3} \zeta_3^2
          - \frac{31256}{27} \zeta_2
          + \frac{208256}{9} \zeta_2 \zeta_3
          - \frac{63496}{45} \zeta_2^2
          - \frac{257344}{63} \zeta_2^3
          \bigg\}
       + C_F^3 n_f\bigg\{
           4  b^q_{4,n_f C_F^3}    
          \bigg\}

       + C_F^3 C_A       \bigg\{
          - \frac{25856}{27}
          + 274432 \zeta_5
          - \frac{18112}{3} \zeta_3
          - \frac{511840}{3} \zeta_3^2
          - \frac{78080}{27} \zeta_2
          - 73728 \zeta_2 \zeta_5
          - \frac{1275328}{9} \zeta_2 \zeta_3
          + \frac{478336}{45} \zeta_2^2
          + 30400 \zeta_2^2 \zeta_3
          + \frac{406912}{15} \zeta_2^3
          \bigg\}

       + C_F^4    \bigg\{
           983040 \zeta_7
          - 49152 \zeta_5
          + 4096 \zeta_3
          - 15360 \zeta_3^2
          - 491520 \zeta_2 \zeta_5
          + 32768 \zeta_2 \zeta_3
          - \frac{391168}{5} \zeta_2^2 \zeta_3
          \bigg\}\,,  
\end{autobreak}  
\end{align}
%
where the unknown coefficients are given \cite{Das:2019btv} numerically $b^q_{4,d_F^{abcd}d_F^{abcd}}=-143.6, b^q_{4,n_f C_F^3}=80.780, b^q_{4,n_f C_F^2 C_A}=-455.247, f^q_{4, d_F^{abcd} d_A^{abcd}}=-100\pm 100$. In the above expression,  $n_f$ is the number of light quarks and the quartic color factors are given by 
\begin{align}
 \frac{d_{F}^{abcd}d_{A}^{abcd}}{N_{F}} =\frac{1}{48} (n_c^2-1)(n_c^2+6) \,,   
 \frac{d_{F}^{abcd}d_{F}^{abcd}}{N_{F}} = \frac{1}{96 n_c^3}(n_c^2-1)(n_c^4-6n_c^2+18)\,,
\end{align}
with $N_F\equiv n_c=3$ in QCD. Although the complete $b\bar{b} H$ form factor is now known to 4-loop \cite{Chakraborty:2022yan}, to obtain the complete $\delta(\zb)$ coefficients, one needs additionally the finite part of the four-loop soft function which is currently not available. We have also computed coefficients for the first four NSV logarithms i.e.\ ${\color{black} \pmb{ L_k }} \equiv \ln^k \zb$ for $k=7,6,5,4$ at fourth order in QCD,
\begin{align}
    
\begin{autobreak}
           \Delta_{\calX, b \bar b}^{\rm NSV, {(4)}} =  

   {\color{black} \pmb{ L_7 }} \bm{ \bigg\{ }C_F^4   \bigg(  - \frac{4096}{3} \bigg) \bm{ \bigg\} } 
   +  {\color{black} \pmb{ L_6 }} \bm{ \bigg\{ } C_F^3 C_A   \bigg(
           \frac{39424}{9}
          \bigg)

       + C_F^4   \bigg(
           \frac{19712}{3}
          \bigg)

       + n_f C_F^3   \bigg(
          - \frac{7168}{9}
          \bigg) \bm{ \bigg\} }
         
          + {\color{black} \pmb{ L_5 }} \bm{ \bigg\{ }  C_F^2 C_A^2    \bigg(
          - \frac{123904}{27}
          \bigg)

       + C_F^3 C_A    \bigg(
          - \frac{805376}{27}
          + 3072 \zeta_2
          \bigg)

       + C_F^4   \bigg(
           2944
          + 20480 \zeta_2
          \bigg)

       + n_f C_F^2 C_A  \bigg(
           \frac{45056}{27}
          \bigg)

       + n_f C_F^3    \bigg(
           \frac{139520}{27}
          \bigg)

       + n_f^2 C_F^2    \bigg(
          - \frac{4096}{27}
          \bigg) \bm{ \bigg\} }

          +{\color{black} \pmb{ L_4 }} \bm{ \bigg\{ }    C_F C_A^3    \bigg(
           \frac{42592}{27}
          \bigg)

       + C_F^2 C_A^2   \bigg(
           \frac{976490}{27}
          - \frac{14240}{3} \zeta_2
          \bigg)

       + C_F^3 C_A    \bigg(
           \frac{1159424}{27}
          - \frac{571840}{9} \zeta_2
          - 8960 \zeta_3
          \bigg)

       + C_F^4   \bigg(
          - \frac{20480}{3}
          - 71168 \zeta_2
          - \frac{286720}{3} \zeta_3
          \bigg)

       + n_f C_F C_A^2    \bigg(
          - \frac{7744}{9}
          \bigg)

       + n_f C_F^2 C_A    \bigg(
          - \frac{344728}{27}
          + \frac{2560}{3} \zeta_2
          \bigg)
       + n_f C_F^3   \bigg(
          - \frac{186752}{27}
          + \frac{87040}{9} \zeta_2
          \bigg)

       + n_f^2 C_F C_A    \bigg(
           \frac{1408}{9}
          \bigg)

       + n_f^2 C_F^2    \bigg(
           \frac{29288}{27}
          \bigg)

       + n_f^3 C_F   \bigg(
          - \frac{256}{27}
          \bigg)
  \bm{ \bigg\} } + \mathcal{O}({\color{black} \pmb{ L_3 }}) \,.
\end{autobreak}
\end{align}%
The highest three logarithms are already computed in \cite{AH:2020iki}, however, the coefficient $\ln^4\zb$ term is a new result. The computation of $\ln^4\zb$ coefficient in the above expression is possible due to the maximally non-abelian property for the diagonal NSV coefficients $\varphi_{I,i}^{(i)}$ \cite{Grunberg:2009vs,AH:2020iki}.\footnote{We thank S. Moch for pointing this out.} Indeed, the diagonal coefficients can be given as $\varphi_{I,l}^{(l)}=-16/l^2 \beta_0^{l-2}C_I^2$, with $C_I=C_F$ for DY and $C_I=C_A$ for Higgs production in gluon fusion. This observation  leads to the following form for the $\varphi_{b,4}^{(4)}$ coefficient,
\begin{align}
\varphi_{b,4}^{(4)} = -C_A^2 C_F^2 \bigg\{ \frac{121}{9} \bigg\} +
 C_A n_f C_F^2 \bigg\{ \frac{44}{9}\bigg\} - C_F^2 n_f^2 \bigg\{  \frac{4}{9}\bigg\}\,.
\end{align}
%%
%%%%%%%%%%%%%%%%%%%%%%%%%%%%%%%%%%%%%
\subsection{Rapidity distribution}
%%%%%%%%%%%%%%%%%%%%%%%%%%%%%%%%%%%%
In the following, we discuss the computation of the SV+NSV rapidity-dependent partonic coefficient function denoted by $\Delta^{b}_{d}$ as given in \eq{eq:SV-NSV}. The procedure follows similar to the inclusive cross-section case described above. In \cite{AH:2020qoa}, $\Delta^{b}_{d}$ was found to be exponentiated in terms of the finite function $\Psi_d^b$ as follows
\begin{align}\label{delta}
\Delta^{b}_{d}(\mu_R^2,\mu_F^2,z_1,z_2) ={\cal C} \exp
\Big({\Psi^b_{d}(\mu_R^2,\mu_F^2,\zo,\zt,\epsilon)}\Big)\, \Big|_{\epsilon = 0} \,.
\end{align}
Similar to \eq{eq:MASTER-XSECT-FAC}, the function $\Psi^b_d$ factorises as,
\begin{align}
\label{eq:Psi-rap}
&\Psi_{d}^b\big(\mu_R^2,\mu_F^2,\zo,\zt,\varepsilon\big) = \bigg( \ln \big( Z_{b} (\ash,\mu^2,\mu_R^2,\varepsilon ) \big)^2   
+ \ln \big| \widehat{F}_{b}\big(\ash,\mu^2,-m_H^2,\varepsilon)\big|^2\bigg) \delta(\zo)\delta(\zt)
\nonumber \\ &
+2 \mathrm{\Phi}_{d}^b\big(\ash,\mu^2,m_H^2,\zo,\zt,\varepsilon\big) 
- \mathcal{C} \ln  \Gamma_{b\bar b}\big(\ash,\mu^2,\mu_F^2,\zo,\varepsilon\big)\delta(\zt) 
- \mathcal{C}\ln \Gamma_{b \bar b}\big(\ash,\mu^2,\mu_F^2,\zt,\varepsilon\big)\delta(\zo) \,.
\end{align}
In contrast to the inclusive case in \eq{eq:MASTER-XSECT-FAC}, the above factorization formula depends on two variables $z_1$ and $z_2$. In addition, it also contains a rapidity-dependent soft function $\Phi_d^b$ which is discussed in detail in \cite{AH:2020qoa}. The NSV improved description of the soft function $\Phi_d^b$ leads to an integral representation of $\Psi_d^b$ as follows,
\begin{align}
\label{eq:psiint}
\Psi^b_{d} =& {1\over 2}\delta(\overline z_1)  \Bigg(\!\!\displaystyle {\int_{\mu_F^2}^{q^2 \overline z_2}
\!\!{\df \lambda^2 \over \lambda^2}}\! {\cal P}_{b \bar b}\left(a_s(\lambda^2),\zt\right) 
\!+\! {\cal Q}^b_{d}\left(a_s(q_2^2),\zt\right)
\!\!\Bigg)_+ 
\nn &
+ {1 \over 4} \Bigg( {1 \over \overline z_1 }
\Bigg\{{\cal P}_{b \bar b}(a_s(q_{12}^2),\zt ) 
+2 C^b(a_s(q_{12}^2)) \ln(\zt)  + 2 D^b(a_s(q_{12}^2))
\nn &
 + q^2{\df \over \df q^2} 
\left({\cal Q}^{b}_{d}(a_s(q_{12}^2 ),\zt) +  {\color{black} 2 }\varphi_{d,b}^f(a_s(q_{12}^2 ),\zt)\right)
\Bigg\}\Bigg)_+
\nn &
+ {1 \over 2}
\delta(\overline z_1) \delta(\overline z_2)
\ln \Big(g^b_{d,0}(a_s(\mu_F^2))\Big)
+ \overline z_1 \leftrightarrow \overline z_2,
\end{align}
where ${\cal P}_{b \bar b} (a_s, \overline z_l)= P_{b \bar b}(a_s, z_l) - 2 B^b(a_s) \delta(\overline z_l)$, $q_l^2 = q^2\bar{z}_l $ and $q^2_{12}=q^2 \overline z_1 \overline z_2$. 
The function ${\cal Q}^b_{d}$ is expressed as,
\begin{align}
{\cal Q}^b_{d}(a_s,\overline z_l) = {2 \over \overline z_l}  \mathbf{D}_{d}^b(a_s) + 2 \varphi^f_{d,b} (a_s,\overline z_l)\,,
\end{align}
where the SV coefficient $ \mathbf{D}_d^b$ are known \cite{Banerjee:2017cfc} to third order in QCD. The function $\varphi_{d,b}^f$ is the finite part of the NSV contribution  to the soft function $\Phi_{d}^{b}$ and has the following expansions,
\begin{align}
\label{eq:Phidf}
\varphi_{d,b}^f(a_s(\mu^2),\overline z_l) 
&= 
\sum_{i=1}^\infty \sum_{k=0}^i a_s^i(\mu^2) \varphi^{b,(k)}_{d,i} \ln^k \overline z_l\,.
\end{align} 
The coefficients $\varphi^{b,(k)}_{d,i}$  are known to the third order in \cite{Ajjath:2020lwb} and are listed below, 
\begin{align}
\label{eq:Phiq}
\varphi^{b,(0)}_{d,1} &= 
        2 C_F \,,
\nonumber \quad 
\varphi^{b,(1)}_{d,1} = 0\,,
\nonumber \\ 
\varphi^{b,(0)}_{d,2} &= C_F n_f   \bigg\{  - \frac{268}{27}         + \frac{8}{3} \zeta_2 \bigg\} + C_F C_A   \bigg\{            \frac{1000}{27} - 28 \zeta_3 - \frac{56}{3} \zeta_2         \bigg\} + C_F^2   \bigg\{  - 16 \zeta_2 \bigg\}\,,
\nonumber\\
\varphi^{b,(1)}_{d,2} &=
        C_F C_A   \bigg\{ 10 \bigg\} + C_F^2   \bigg\{  - 10 \bigg\}\,,
\nonumber \quad
\varphi^{b,(2)}_{d,2} =
        C_F^2   \bigg\{  - 4 \bigg\}\,,
\nonumber\\
\varphi^{b,(0)}_{d,3} &=
       C_F n_f^2   \bigg\{
           \frac{10856}{729}
          + \frac{32}{27} \zeta_3
          - \frac{304}{27} \zeta_2
          \bigg\}

       + C_F C_A n_f   \bigg\{
          - \frac{118984}{729}
          + \frac{196}{3} \zeta_3
          + \frac{11816}{81} \zeta_2 
          \nonumber\\ &
          - \frac{208}{15} \zeta_2^2
          \bigg\}

       + C_F C_A^2   \bigg\{
           \frac{576020}{729}
          + 192 \zeta_5
          - \frac{21692}{27} \zeta_3
          - \frac{40844}{81} \zeta_2
          + \frac{176}{3} \zeta_2 \zeta_3
          + \frac{656}{15} \zeta_2^2
          \bigg\}
   \nonumber\\ &
   
       + C_F^2 n_f   \bigg\{
          - \frac{1144}{9}
          + 96 \zeta_3
          + \frac{1432}{27} \zeta_2
          + \frac{32}{5} \zeta_2^2
          \bigg\}

       + C_F^2 C_A   \bigg\{
          - \frac{4279}{27}
          + \frac{460}{9} \zeta_3
          - \frac{5548}{27} \zeta_2
             \nonumber\\ &
   
          + \frac{1312}{15} \zeta_2^2
          \bigg\}

       + C_F^3   \bigg\{
           23
          + 48 \zeta_3
          - \frac{32}{3} \zeta_2
          - \frac{448}{15} \zeta_2^2
          \bigg\}\,, 
\nonumber\\
\varphi^{b,(1)}_{d,3} &=
            C_F C_A n_f   \bigg\{
          - \frac{256}{9}
          - \frac{28}{9} \zeta_2
          \bigg\}

       + C_F C_A^2   \bigg\{
           \frac{316}{9}
          + 24 \zeta_3
          - \frac{8}{9} \zeta_2
          \bigg\}

       + C_F^2 n_f   \bigg\{
           \frac{3952}{81}
          - \frac{64}{9} \zeta_2
          \bigg\}
   \nonumber\\ &
       + C_F^2 C_A   \bigg\{
          - \frac{19732}{81}
          + \frac{544}{3} \zeta_3
          + \frac{436}{9} \zeta_2
          \bigg\}

       + C_F^3   \bigg\{
          - \frac{64}{3}
          - 64 \zeta_3
          + \frac{80}{3} \zeta_2
          \bigg\}\,,
\nonumber\\
\varphi^{b,(2)}_{d,3} &=
       
       C_F C_A n_f   \bigg\{
          - \frac{10}{3}
          \bigg\}

       + C_F C_A^2   \bigg\{
           34
          - \frac{10}{3} \zeta_2
          \bigg\}

       + C_F^2 n_f   \bigg\{
           \frac{40}{3}
          \bigg\}
 \nonumber\\ &
       + C_F^2 C_A   \bigg\{
          - 96
          + \frac{52}{3} \zeta_2
          \bigg\}
 
       + C_F^3   \bigg\{
           \frac{16}{3}
          \bigg\}\,,

\nonumber\\
\varphi^{b,(3)}_{d,3} &=
        C_F^2 n_f   \bigg\{ \frac{32}{27} \bigg\} + C_F^2 C_A   \bigg\{ - \frac{176}{27} \bigg\}\,.
\end{align}

Similar to the inclusive case, the resummation is performed in the conjugate space, and in particular, one now needs double Mellin transformations (M-M) corresponding to both threshold variables $z_1,z_2$,
\begin{align} \label{eq:DeltaN-rap}
\Delta_{d, \NB_1, \NB_2}^b =  \prod\limits_{i=1,2} \int_0^1 dz_i z_i^{N_i-1} \Delta_d^b(z_1,z_2) 
 \equiv  g_{d,0}^b \exp(\Psi_{d,\NB_1, \NB_2}^b)\,,
\end{align}
where $\NB_i \equiv N_i\exp(\gamma_{E})$ and $ g_{d,0}^b = \sum\limits_{i=0}^{\infty} a_s^i g_{d,0,i}^b$ are the $N$-independent constants which, as before, contain the process-dependent information from the form factor and universal constants arising due to the Mellin transformations. The resummed exponent $\Psi_{d, \NB_1, \NB_2}^b$ in this case takes the following form,
\begin{align}
\label{eq:PsiN}
\Psi_{d, \NB_1, \NB_2}^b = &~~
  g_{d,1}^b(\omega_{12})  \ln \NB_1
+ \sum\limits_{i=0}^\infty a_s^i \bigg( \frac{1}{2}  g_{d,i+2}^b(\omega_{12}) + \frac{1}{N_1} \overline{g}_{d,i+1}^b(\omega_{12}) \bigg)
\nn &
 +\frac{1}{N_1} \left(h^b_{d,0}(\omega_{12},\NB_1)  + 
\sum\limits_{i=1}^{\infty} a_s^i h^b_{d,i}(\omega_{12},\omega_1,\NB_1)\right) 
+ (1 \leftrightarrow 2) \,,
\end{align}
where
\begin{align}
\label{hg}
h^b_{d,0}(\omega_{12},\NB_l) &= h^b_{d,00}(\omega_{12}) \,,
\nn
         h^b_{d,i}(\omega_{12},\omega_l,\NB_l) &= \sum_{k=0}^{i-1} h^b_{d,ik}(\omega_{12})~ \ln^k \NB_l 
         + \tilde{h}^b_{d,ii}(\omega_{12},\omega_l)~ \ln^k \NB_l\,.
\end{align}
Here $\omega_{12} = a_s \beta_0 \ln \NB_1 \NB_2$ and $\omega_l = a_s \beta_0 \ln \NB_l$ for $l=1,2$. The diagonal terms $\tilde{h}^b_{d,ii}$ at each order in QCD are given by
\begin{align}\label{hg-diag}
\tilde{h}^b_{d,11}(\omega_{12},\omega_l) =& h^b_{d,11}(\omega_{12}) + \frac{\omega_l}{\beta_0} h^b_{d,22}
\nonumber \\
\tilde{h}^b_{d,ii}(\omega_{12},\omega_l) = & \frac{\omega_l}{\beta_0} h^b_{d,i+1,i+1}\,, ~ \forall ~i \ge 2\,.
\end{align}
Similar to the inclusive case, the entire all-order information is embedded systematically in the resummation exponents $g^b_{d,0,i},g^b_{d,i}(\omega),\overline {g}^b_{d,i}(\omega)$ and $h^b_{d,i}(\omega,\omega_l,N_l)$, which can be used to predict SV and NSV terms to all orders. The resummation coefficients ${{g}^b_{d,0,0}, g^b_{d,1}}$ are sufficient to predict the leading SV terms $a_s^i \ln^l \NB_1 \ln^k \NB_2$ with $l+k=2i$ (where $l, k \geq 0$) for all $i > 1$. These terms form the tower of leading logarithms (LL) in SV resummation. Including the functions ${{g}^b_{d,0,1}, g^b_{d,2}}$ along with the previous set allows for the prediction of two towers of next-to-leading logarithmic (NLL) SV terms, expressed as $a_s^i \ln^l \NB_1 \ln^k \NB_2$ with $l+k=2i-1, 2i-2$ for all $i > 2$. Using both the first and second sets of resummation coefficients, one can derive the logarithms contributing to the SV-NLL resummation. In general, the resummed results, characterized by the exponents ${{g}^b_{d,0,n}, g^b_{d,n+1}}$ along with the preceding sets can predict terms of the form $a_s^i \ln^l \NB_1 \ln^k \NB_2$ with $l+k=2n+1, 2n$ for all $i > n+1$, where $n = 0, 1, 2, \ldots$, contributing to the SV-${\rm{N^nLL}}$ resummation. Up to the third order, all these coefficients can be found in \cite{Das:2023bfi}. For NSV resummation, using the first set of resummation coefficients ${{g}^b_{d,0,0}, g^b_{d,1}, \overline{g}^b_{d,1}, h^b_{d,0}}$, one can predict the leading towers of NSV logarithms ${a_s^i \frac{\ln^l \NB_1}{N_1} \ln^k \NB_2, a_s^i \frac{\ln^l \NB_2}{N_2} \ln^k \NB_1}$ with $l+k=2i-1$ for all $i>1$. These leading towers contribute to the NSV-$\overline{\rm LL}$ resummation. Incorporating the second set of resummation coefficients ${{g}^b_{d,0,1}, g^b_{d,2}, \overline{g}^b_{d,2}, h^b_{d,1}}$ along with the first set allows for the prediction of the towers of next-to-leading NSV terms ${a_s^i \frac{\ln^l \NB_1}{N_1} \ln^k \NB_2, a_s^i \frac{\ln^l \NB_2}{N_2} \ln^k \NB_1}$ with $l+k=2i-2$ for all $i>2$. These towers contribute to the NSV-$\overline{\rm NLL}$ resummation. In general, by employing the $n$-th set ${{g}^b_{d,0,n}, g^b_{d,n+1}, \overline{g}^b_{d,n+1}, h^b_{d,n}}$ in addition to the previous sets, one can predict the highest $(n+1)$ towers of NSV logarithms in $\NB_l$ (with $l=1,2$). These contribute to the $\overline{\rm N\text{\it n}LL}$ resummation, covering every order in $a_s^i$ for all $i>n+1$. All these NSV resummed exponents up to the third order are collected in Appendix\ \app{App:NSV-RAPIDITY-COEFF}. 

Finally, one is required to complement the resummed results with the missing regular contributions from the fixed order. This is achieved through a matching procedure that follows a similar footstep as the inclusive cross-section and takes the form,
\begin{align}\label{eq:matchrap}
\frac{\df \sigma^{\rm {N{\it n}LO+\overline {\rm N{\it n}LL}}}(\tau,y)}{\df y}
=&  
{\df\sigma(\tau, y)\over \df y } \bigg|_{\rm N{\it n}LO}
+
\,  \sigma_{b \bar{b}}^{(0)}(\mu_R) 
\sum_{ij\in\{b,\bar{b}\}} \int_{c_{1} - i\infty}^{c_1 + i\infty} \frac{\df N_{1}}{2\pi i}
 \int_{c_{2} - i\infty}^{c_2 + i\infty} \frac{\df N_{2}}{2\pi i} 
{\tau}^{-N_{1}-N_{2}} 
\nn
&\times
f_{i,N_1}(\mu_F^2)  f_{j,N_2}(\mu_F^2) 
\bigg( \Delta_{d,\NB_1,\NB_2}^b \big|_{\overline {\rm {N{\it n}LL}}} - {\Delta^b_{d,\NB_1,\NB_2}}\big|_{\rm {tr{\text -}N{\it n}LO}}  \bigg) \,.
\end{align}
The matching procedure avoids double counting of SV and NSV logarithms by removing them using the last term ${\Delta^b_{d,\NB_1,\NB_2}}\big|_{\rm {tr{\text -}N{\it n}LO}}$.
%%%%%%%%%%%%%%%%%%%%%%%%%%%%%%%%%%%%%%%%%%%%%%%%
\section{Numerical Results}\label{sec:NUMERICS}
%%%%%%%%%%%%%%%%%%%%%%%%%%%%%%%%%%%%%%%%%%%%%%%%
We now have all the ingredients to study the effect of SV and NSV resummations for bottom-induced Higgs production at the LHC for $13$ TeV center-of-mass energy. We fix the central scale at $(\mur^c,\muf^c ) = (1,1/4) m_h $, as this choice is known \cite{Maltoni:2003pn} to reduce large collinear effects. Our default choice of PDF is the \textsc{CT14} \cite{Dulat:2015mca} set, which is implemented through the \textsc{LHAPDF6} \cite{Buckley:2014ana} interface.\footnote{The choice of this set is due to the availability of fixed-order N2LO rapidity distribution \cite{Mondini:2021nck} with this PDF. However, as stated in the next sections, this is not a constraint for the total cross-section or the resummed results.} The strong coupling is evolved order-by-order using the QCD beta function \cite{Gross:1973ju,Politzer:1973fx,Caswell:1974gg,Jones:1974eu,Egorian:1978zx,Tarasov:1980au,Larin:1993tp,vanRitbergen:1997va,Czakon:2004bu,Baikov:2016tgj,Herzog:2017ohr,Luthe:2017ttg} and we set the initial value $\alpha_s(m_Z=91.1876 \text{ GeV}) = 0.118$ beyond NLO level. The Yukawa coupling is evolved using the mass anomalous dimension with bottom mass $m_b(m_b) = 4.18$ GeV in the $\overline{\text{MS}}$ scheme. Up to four loops, these mass anomalous dimensions are available in \cite{Tarasov:1982plg,Larin:1993tq,Alekseev:1994xp,Chetyrkin:1997dh,Czakon:2004bu}. The five-loop quark mass anomalous dimension needed at the N4LO level is obtained from \cite{Baikov:2014qja,Baikov:2017ujl}. The RGE for the Yukawa in terms of the mass anomalous dimensions $\gamma_m$ takes the following form,
\begin{align}
        \mur^2 \frac{\df}{\df \mur^2} \lam_y(\mur) 
        =
        \gamma_m(a_s) \lam_y(\mur)\,,
\end{align}
with $\gamma_m= -\sum_{i=0}^{\infty} \as^{i+1} \gamma_{m}^{(i)}$. The numerical values of $\gamma_m^{(i)}$ in QCD up to the required orders are as follows,
\begin{align}
\gamma_{m}^{(0)} &= 4\,, 
\nn
\gamma_{m}^{(1)} &= 67.333 - 2.222 ~\NF\,, 
\nn
\gamma_{m}^{(2)} &= 1249.0 - 146.184 ~\NF - 1.728 ~\NF^2\,, 
\nn
\gamma_{m}^{(3)} &= 25329.514 - 4891.510 ~\NF + 70.698 ~\NF^2 + 1.483 ~\NF^3 \,,
\nn
\gamma_{m}^{(4)} &= 573139.861 - 147134.942 ~\NF  + 7661.955 ~\NF^2 + 110.918 ~\NF^3 - 0.0874 ~\NF^4\,,
\end{align}
with $n_f=5$ as the number of light quark flavors. For the choice of contour in the Mellin inversion in \eq{eq:matchincl} and \eq{eq:matchrap} we follow the procedure described in \cite{AH:2019phz,Das:2023rif} and set $c=c_1=c_2=1.9$ to avoid the Landau pole. 

For the phenomenological studies, we used the fixed order inclusive cross-section up to N3LO from the publicly available code {\tt n3loxs} \cite{Baglio:2022wzu}. On the other hand, the fixed order rapidity distribution is available to N2LO from \cite{Mondini:2021nck} which uses the N-Jettiness \cite{Boughezal:2015dva,Gaunt:2015pea} slicing method as implemented in a private version of MCFM \cite{Campbell:1999ah,Campbell:2015qma,Boughezal:2016wmq}. All the resummed results were obtained using an in-house code with the same set of parameters as the fixed order.

We define the following ratios to estimate the impact of SV(NSV) resummation over the fixed order (for both total cross-section and rapidity), 
\begin{align}\label{eq:RF-factor}
 RF_{ij} = \frac{\sigma_{\rm N{\it i}LO+N{\it i}LL}}{\sigma_{\rm N{\it j}LO}} \,,
\overline{RF}_{ij} = \frac{\sigma_{\rm N{\it i}LO+\overline{N{\it i}LL}}}{\sigma_{\rm N{\it j}LO}} \,.
\end{align}
We also define the following quantities to estimate the perturbative convergence,
\begin{align}\label{eq:K-R-factor}
 K_{ij} = \frac{\sigma_{\rm N{\it i}LO}}{\sigma_{\rm N{\it j}LO}} \,,
{R}_{ij} = \frac{\sigma_{\rm N{\it i}LO+{N{\it i}LL}}}{\sigma_{\rm N{\it j}LO+{N{\it j}LL}}}\,,
\overline{R}_{ij} = \frac{\sigma_{\rm N{\it i}LO+\overline{N{\it i}LL}}}{\sigma_{\rm N{\it j}LO+\overline{N{\it j}LL}}} \,.
\end{align}
%%
%%%%%%%%%%%%%%%%%%%%%%%%%%%%%%%%%%%%%%%%%%%%%%%%%%%%%%%%%%%%%%%%%%
\subsection{Inlcusive production}\label{sec:NUMERICS-INCLUSIVE}
\begin{figure}[ht!]
\centerline{
\includegraphics[width=12.5cm,height=10.0cm]{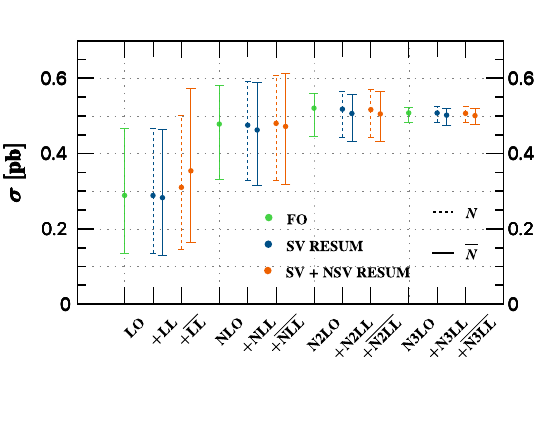}}
\caption{\small{The behavior of cross-sections at FO, SV-resummed, and SV+NSV-resummed is shown in different orders along with the seven-point scale uncertainties. For the resummed results, the two different schemes $N$ (dashed) and $\overline{N}$ (solid) are shown.}}
\label{fig:xsect-n-vs-nb}
\end{figure}

%%%%%%%%%%%%%%%%%%%%%%%%%%%%%%%%%%%%%%%%%%%%%%%%%%%%%%%%%%%%%%%%%%
We begin by examining the total inclusive cross-section. In \fig{fig:xsect-n-vs-nb}, we compare the fixed order calculation with SV and SV+NSV resummations up to the third order in QCD. Additionally, we compare two different schemes for the resummation in the SV and SV+NSV cases viz.\ the {\it $N$-scheme} (dashed line) and the {\it $\overline{N}$-scheme} (solid line) as outlined in \sect{sec:incl-xsect}. In general, both SV and SV+NSV resummations show better perturbative convergence than fixed-order calculations, particularly at the first few lower orders. However, at higher orders, the differences between these approaches diminish. When comparing the {\it $N$-scheme}  and {\it $\overline{N}$-scheme}, we find that the latter converges more quickly—a behavior also observed in DIS \cite{Das:2019btv} and DY \cite{AH:2020cok} processes for the SV resummation case.  We observe a similar trend for SV+NSV resummation as well.

Regarding the seven-point scale uncertainty, we notice that it does not improve over the fixed-order results for either SV or SV+NSV resummations; however, the central values in both cases become more stable.
%%%
%%%
Quantitatively, at fixed order, we find that NLO introduces $65\%$ correction over LO, N2LO adds $9\%$ over NLO, and N3LO yields a correction of $-2.5\%$ over N2LO. When examining the truncated cross-sections at SV and SV+NSV approximations in both momentum-space and Mellin-space, we observe that the momentum-space SV+NSV captures the perturbative behavior of the fixed order slightly better overall. However, this does not necessarily imply that it is preferable for resummation. In fact, at the N2LO level, the momentum-space SV+NSV result closely approximates the fixed-order result. In contrast, at N3LO, the truncated SV result in Mellin-space aligns better with the fixed-order results. In \tab{tab:n3lo-xs} we quote both the fixed order (N3LO) and the resummed predictions (N3LL and $\overline{\rm N3LL}$)\footnote{We use shorthand notations +N${\it n}$LL and +$\overline{\rm N{\it n}LL}$ to denote the matched results where resummed results are matched with the fixed order at the same level.} along with their corresponding seven-point scale uncertainties. The last two columns stand for the resummed predictions in the {\it $\overline{N}$-scheme}. At higher orders, the differences among different scheme choices are reduced as reflected from \tab{tab:n3lo-xs}.
%%%%
\begin{table}[ht!]
\centering{
\begin{tabular}{|P{1.8cm}|P{2.3cm}|P{2.3cm}||P{2.3cm}|P{2.3cm}|}
    \hline 
    \multicolumn{1}{|c|}{N3LO} &
    \multicolumn{1}{|c|}{+N3LL ($N$)} &
    \multicolumn{1}{c||}{+$\overline{\rm N3LL}$ ($N$)} &
    \multicolumn{1}{c|}{+{N3LL} ($\overline{N}$)} & 
    \multicolumn{1}{c|}{+$\overline{\rm N3LL}$  ($\overline{N}$)} \\ 
    \hline
    \hline
   0.508(25) & 0.506(25) & 0.505(24) & 0.503(26) & 0.502(24)  \\ 
    \hline
  \end{tabular}
  }
  \caption{The third order cross-section at fixed order with SV and SV+NSV resummation along with different 
  scheme choices.}
    \label{tab:n3lo-xs}
\end{table}
%%%%
For the rest of the article, we fix the {\it $\overline{N}$-scheme} as our default choice. 

%%%
\begin{figure}[ht!]
\centerline{
\includegraphics[width=7.5cm,height=5.0cm]{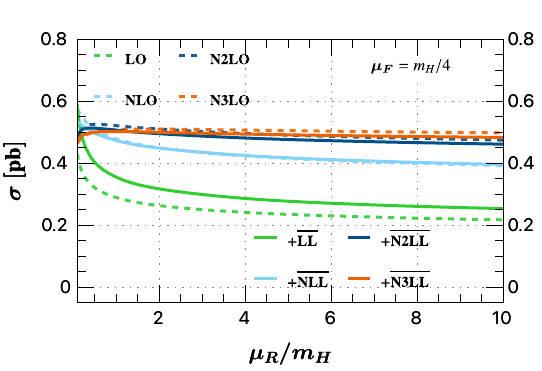}
\includegraphics[width=7.5cm,height=5.0cm]{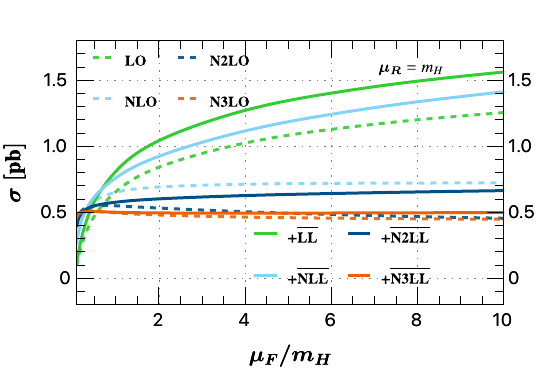}}
\caption{\small{The behavior of the $\mu_R$ (left) and $\mu_F$ (right) variations at fixed order (dashed) and SV+NSV resummed order (solid) are presented.}}
\label{fig:xsect-scalevar-nsv}
\end{figure}
%%%
In \fig{fig:xsect-scalevar-nsv}, we study the dependence of the 
renormalization scale ($\mur$) while fixing the factorization scale at its central value, and vice-versa, for both fixed order and SV+NSV resummation in {\it $\overline{N}$-scheme}. For this purpose, we vary the scale in the range $[1/10,10]m_H$. At the N3LO level, we observe that $\mur$ scale uncertainty improves to $5.1\%$ from $8.4\%$ by the inclusion of $\overline{\rm N3LL}$ resummation. The uncertainty is computed from the absolute maximum deviation from the central scale choice. A similar method yields an improvement for the $\muf$ uncertainty to $8.2\%$ from $12.2\%$ uncertainty of the fixed order. However, the $\muf$ variation gets doubled compared to fixed-order considering the end values. This is a known behavior, and to improve the $\muf$ scale uncertainty, the resummation of NSV logarithms from off-diagonal channels is also needed.
%%%
\begin{figure}[ht!]
\centerline{
\includegraphics[width=7.5cm,height=5.0cm]{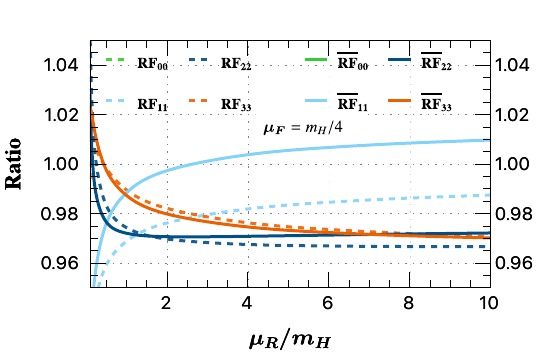}
\includegraphics[width=7.5cm,height=5.0cm]{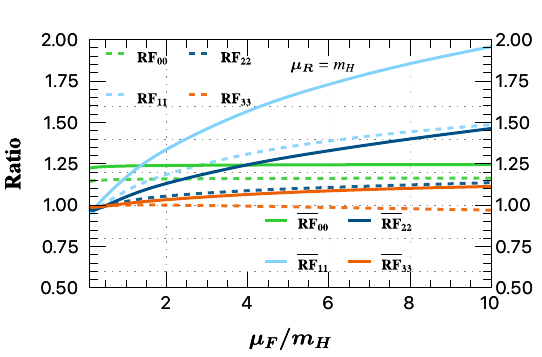}}
\caption{\small{Same as \fig{fig:xsect-scalevar-nsv} but for the $RF$ and $\overline{RF}$ ratios defined in \eq{eq:RF-factor}.}}
\label{fig:xsect-ratiovar-nsv}
\end{figure}
%%%
In \fig{fig:xsect-ratiovar-nsv} we compare the effects of SV+NSV resummation against the SV resummation through the $RF_{ij}$ and $\overline{RF}_{ij}$ ratios defined in \eq{eq:RF-factor}. At higher orders, the SV+NSV resummed results show better stability over the SV resummed results for the $\mur$ variation. On the other hand, the $\muf$ scale uncertainty is better for the SV resummation. This again reflects the importance of including the NSV resummed corrections from the off-diagonal channels \cite{Lustermans:2019cau,vanBeekveld:2021mxn}.

%%%%
\begin{table}[ht!]
\centering{
\begin{tabular}{|P{1.5cm}|P{1.6cm}|P{1.6cm}|P{1.6cm}|P{1.6cm}|P{1.6cm}|P{1.6cm}|}
    \hline 
    \multicolumn{1}{|c|}{$E_{\rm CM}$(TeV)} &
    \multicolumn{1}{c|}{N2LO} &
    \multicolumn{1}{c|}{+N2LL} &
    \multicolumn{1}{c|}{+\nsv{N2LL}} & 
    \multicolumn{1}{c|}{N3LO} &
    \multicolumn{1}{c|}{+N3LL} &
    \multicolumn{1}{c|}{+\nsv{N3LL}} \\ 
    \hline
    \hline
    $7$  
    &$0.169(19)$& $0.165(20)$& $0.164(20)$ %% N2LO
    & $0.163(6)$& $0.161(6)$& $0.161(7)$\\   %% N3LO
    \hline
    $8$  
    &$0.219(26)$& $0.213(27)$  & $0.213(26)$ %% N2LO
    & $0.212(8)$& $0.209(8)$& $0.209(9)$\\ %% N3LO
    \hline
    $13$  
    &$0.521(74)$ & $0.507(74)$  & $0.506(73)$ %% N2LO
    & $0.508(25)$&$0.503(26)$& $0.502(24)$\\ %% N3LO
    \hline
    $14$  
    &$0.589(86)$& $0.574(85)$  & $0.572(84)$ %% N2LO
    & $0.576(30)$& $0.570(31)$& $0.568(29)$\\ %% N3LO
    \hline
    $27$  
    &$1.614(281)$& $1.575(278)$  & $1.571(274)$ %% N2LO
    & $1.598(112)$& $1.583(114)$& $1.580(108)$\\ %% N3LO
    \hline
 %   $100$  
 %   &$8.666(1967)$& $8.481(1941)$ & $8.451(1914)$ %% N2LO
 %   & $8.785(934)$& $8.714(936)$& $8.697(907)$\\ %% N3LO
 %   \hline
  \end{tabular}
  }
  \caption{Cross-sections along with the absolute errors obtained from seven-point scale variation at fixed order and SV and SV+NSV resummed orders are shown for different $E_{\rm CM} $ choices.}
    \label{tab:ecm-xsect} 
\end{table}
%%%%
In \tab{tab:ecm-xsect}, we present the inclusive cross-section for various center-of-mass energies, comparing both fixed-order and resummed predictions. At higher $E_{\rm CM}$, the seven-point scale uncertainty shows a marginal improvement over the fixed-order result, with the lowest uncertainty observed for the NSV resummation, amounting to less than $7\%$.
%%%%
\begin{table}[ht!]
\centering{
\begin{tabular}{|P{3.5cm}|P{2.3cm}|P{2.3cm}|}
    \hline 
    \multicolumn{1}{|c|}{PDF set} &
    \multicolumn{1}{c||}{N3LO(pb)} &
    \multicolumn{1}{c|}{+\nsv{N3LL}(pb)} \\ 
    \hline
    \hline
    {\texttt{ABMP16} \cite{Alekhin:2017kpj}}&  $0.5750^{+0.0206}_{-0.0206}$ & $0.5684^{+0.0203}_{-0.0203}$ \\
    \hline
    {\texttt{MSHT} \cite{Bailey:2020ooq}}& $0.5014^{+0.0083}_{-0.0085}$ & $0.4950^{+0.0082}_{-0.0085}$ \\
    \hline
    {\texttt{CT14} \cite{Dulat:2015mca}} & $0.5081^{+0.0233}_{-0.0248}$ & $0.5016^{+0.0230}_{-0.0245}$\\
    \hline
    {\texttt{CT18} \cite{Hou:2019qau}} & $0.5038^{+0.0199}_{-0.0224}$ & $0.4974^{+0.0196}_{-0.0221}$\\
    \hline
    {\texttt{NNPDF40} \cite{NNPDF:2021njg}} & $0.4814^{+0.0037}_{-0.0037}$ & $0.4753^{+0.0036}_{-0.0036}$\\
    \hline
    {\texttt{NNPDF40aN3LO} \cite{NNPDF:2024nan}} & $0.5041^{+0.0036}_{-0.0036}$ & $0.4977^{+0.0035}_{-0.0035}$\\
    \hline
  \end{tabular}
  }
 \caption{The intrinsic PDF uncertainties are shown for both N3LO and NSV resummation at \nsv{N3LL}.}
    \label{tab:pdfuncer-xsect}
\end{table}
%%%%
In \tab{tab:pdfuncer-xsect}, we present the central values along with the intrinsic PDF uncertainties for various PDF sets. We also observe a marginal improvement in PDF uncertainties with SV+NSV resummed results. The central values decrease by up to $1\%$ in the SV+NSV resummed case. The intrinsic PDF uncertainty is the lowest for \texttt{NNPDF40}, at below $1\%$, while it is around $5\%$ for \texttt{CT14} and \texttt{CT18}. The use of approximate N3LO PDF \texttt{NNPDF40aN3LO} \cite{NNPDF:2024nan} shifts the central value by almost $4.7\%$ compared to using the N2LO PDF with the error similar to the N2LO one for both the fixed order as well as the resummed case.

Finally, we estimate the effect of SV correction at the fourth order in perturbative QCD. We formulate the SV correction in the Mellin space from the truncated N4LL resummation formula. Although the exponent is known up to the N4LL level, the finite $N$-independent coefficient ($g_{0,4}^b$) is not completely known due to the missing finite term for the four-loop soft function. Hence for the $g_{0,4}^b$ coefficient we use a [1/1] Pade approximate value $g_{0,4}^b=91358.7$. Compared to the N3LO fixed order, the cross-section decreases by $-0.1\%$ with the scale uncertainty remaining at the same $2.5\%$ as the N3LO. On the other hand, the N4LL SV resummation matched with N4LOsv results gets a decrement of $-0.5\%$ compared to N3LO with the seven-point scale uncertainty remaining around $2.5\%$ showing a stable perturbative convergence.
%%%%%%%%%%%%%%%%%%%%%%%%%%%%%%%%%%%%%%%%%%%%%%%%%%%%%%%%%%%%%%%%
\subsection{Rapidity distribution}\label{sec:NUMERICS-RAPIDITY}
\begin{figure}[ht!]
\centerline{
\includegraphics[width=7.5cm,height=5.0cm]{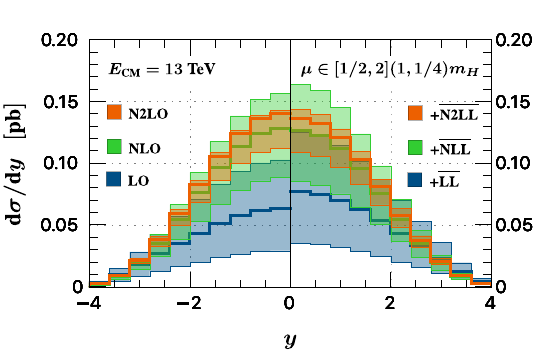}
\includegraphics[width=7.5cm,height=5.0cm]{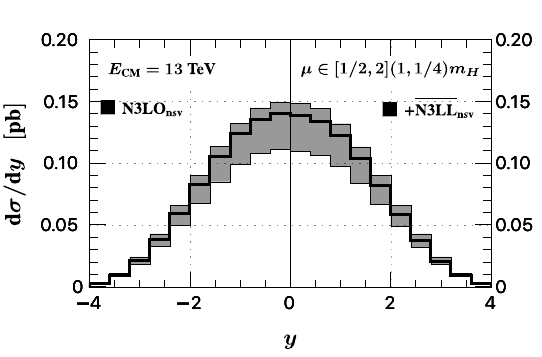}}
\caption{\small{The behavior of rapidity distribution at fixed order and NSV resummed orders along with the corresponding seven-point scale uncertainties. The symbol $+\overline{\rm N{\it n}LL}$ means the NSV resummed result was matched with the corresponding fixed order. On the right panel, the matching is performed using third-order NSV data as the fixed order.}}
\label{fig:rap-scalevar}
\end{figure}

%%%%%%%%%%%%%%%%%%%%%%%%%%%%%%%%%%%%%%%%%%%%%%%%%%%%%%%%%%%%%%%%
Next, we study the effects of NSV resummation on the rapidity distribution. We first compare the SV+NSV resummed results, matched with the available fixed-order results, against the fixed-order results themselves, as shown in \fig{fig:rap-scalevar}. The results up to N2LO and +$\overline{\rm N2LL}$ are shown in the left panel. We observe better perturbative convergence with NSV resummation compared to fixed-order results, similar to the findings for SV resummation in \cite{Das:2023rif}. However, the seven-point scale uncertainty does not improve over the fixed-order results, a behavior also noted for SV resummation. On the right panel of \fig{fig:rap-scalevar}, we compare the third order cross-section at SV+NSV denoted as N3LOnsv with the $\overline{\rm N3LL}$ resummation at the same accuracy ($\overline{\rm N3LL}$nsv). To construct the fixed-order SV+NSV result at the third order, we combined the SV+NSV correction at the third order with the full N2LO result, both calculated using the same parameter sets. Specifically, the strong coupling is evolved using the four-loop beta function. The N3LOnsv gets a negative correction which is $-0.3\%$ at $y=0$ compared to N2LO, whereas at the high rapidity ($y=3.6$), the correction is around $0.01\%$. The resummed $\overline{\rm N3LL}$nsv gets a flat correction over the $+\overline{\rm N2LL}$ by $1.3\%$. The subscript `nsv' implies that in the matching procedure, the N3LOnsv result has been used.
%%%
%%%

\begin{figure}[ht!]
\centerline{
\includegraphics[width=7.5cm,height=5.0cm]{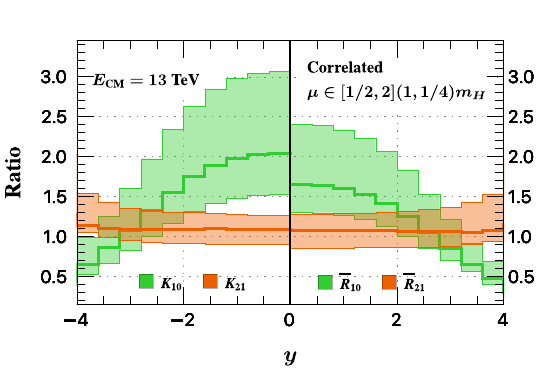}
\includegraphics[width=7.5cm,height=5.0cm]{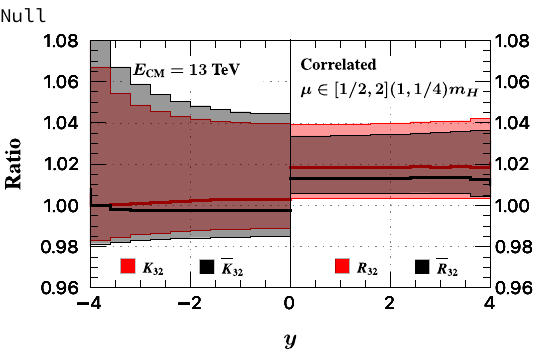}}
\caption{\small{(Left) The $K, R$ factors along with correlated errors are shown up to the second order. (Right) The same is shown in the third order along with the SV results. The third-order results are up to NSV(black) and SV(red) accuracy and the same for matched results.
}}
\label{fig:rap-kr-fac}
\end{figure}

\begin{figure}[ht!]
\centerline{
\includegraphics[width=7.5cm,height=5.0cm]{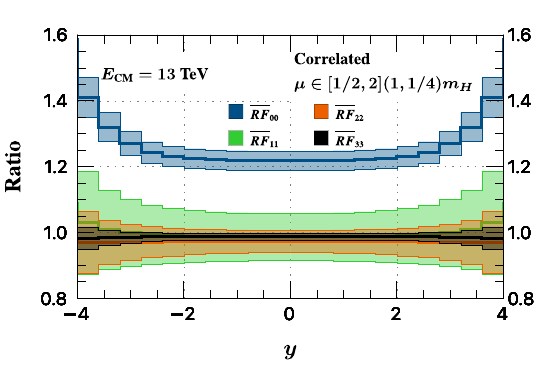}}
\caption{\small{The $\overline{RF}$ factor along with correlated errors are shown up to the third order. For the third order, resummed $\overline{\rm N3LL}$  results are matched to the $ \rm N3LOnsv$ results.}}
\label{fig:rap-rf-fac}
\end{figure}

To observe the effects of higher order correction including the NSV resummation, we analyzed the ratios $K_{ij}$ and $\overline{R}_{ij}$ as defined in \eqref{eq:K-R-factor} in \fig{fig:rap-kr-fac} and \fig{fig:rap-rf-fac}. The correlated scale uncertainties are also shown through the bands. From the left panel of \fig{fig:rap-kr-fac} a better perturbative convergence in the SV+NSV resummed series is seen. On the right panel of \fig{fig:rap-kr-fac} we present the same quantities in the third order along with the SV resummed results. Note that for the $K_{32}$ factor, we have used the third-order SV result whereas for the $\overline{K}_{32}$ factor we have used the third-order SV+NSV result. Similarly for the $R_{32}$ and $\overline{R}_{32}$ ratios, we have matched the N3LL and $\overline{\rm N3LL}$  resummed results with the corresponding N3LOsv and N3LOnsv results. Although we do not observe a scale reduction for N3LOnsv in comparison to the N3LOsv, the resummed results indeed show a scale reduction for $\overline{\rm N3LL}$ throughout the entire rapidity spectrum, along with a central value closer to the N2LO+$\overline{\rm N2LL}$ result showing a better perturbative convergence. 
%%%%
%%%
%%%
%%%
The relative size of the NSV resummed effect is shown over the respective fixed order in  \fig{fig:rap-rf-fac} where the $RF$ ratios as defined in \eq{eq:RF-factor} are shown up to the third order. The NSV resummed results particularly amount to higher cross-section at higher rapidity up to the second order. On the other hand at N2LO and N3LO levels, the corrections are relatively flat over the rapidity region. As before, the fixed order at N3LO is constructed to have up to NSV accuracy and is missing the sub-dominant regular contributions as well as off-diagonal contributions. Nevertheless, we observe a nice perturbative convergence and correlated scale reduction which amounts to around $1\%$ at the third order.

%%%
\begin{figure}[ht!]
\centerline{
\includegraphics[width=7.5cm,height=5.0cm]{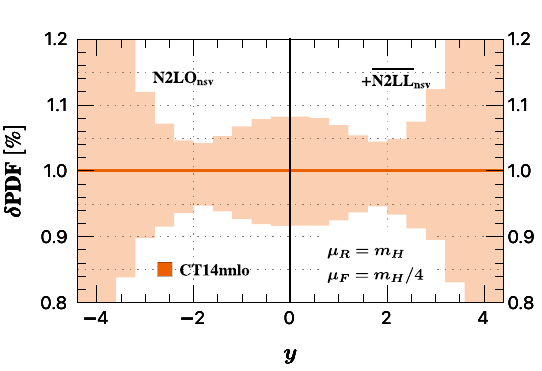}
\includegraphics[width=7.5cm,height=5.0cm]{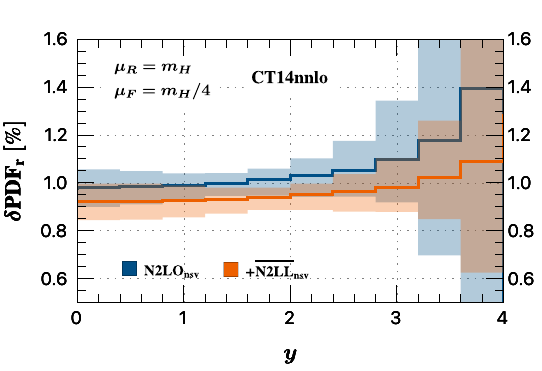}}
\caption{\small{ (Left) Effects of N2LOnsv and matched \nsv{N2LL}  on PDF uncertainty are shown. The subscript `nsv' on the latter refers that it has been matched to the corresponding fixed order with NSV accuracy. (Right) The same with overall N2LO normalization is shown. The solid lines correspond to the results for the normalized 0-th subset in each case.}}
\label{fig:rap-pdf}
\end{figure}
%%%
Finally, in \fig{fig:rap-pdf}, we studied the sensitivity of different PDF subsets on the NSV result from both at the fixed order (N2LOnsv) and the SV+NSV resummation level ($\overline{\rm N2LL}{\rm nsv}$). For this purpose, we have constructed the following ratio,
\begin{align}\label{eq:rap-pdfuncer}
    \delta\text{PDF}= 1 \pm \delta \left[\frac{\df \sigma}{\df y}\right]\bigg/ \left[\frac{\df \sigma}{\df y}\right]_0 \times 100\% \,,
\end{align}
where $\delta \left[\frac{\df \sigma}{\df y}\right]$ is the absolute intrinsic PDF uncertainty calculated through the LHAPDF interface from all the subsets of the PDF. The subscript `$0$' indicates the corresponding quantity is calculated with the central PDF set. At $y=0$ both the fixed order SV+NSV and resummed SV+NSV results amount to around $8\%$ uncertainty which gets reduced to around $4.5\%$ at $y=2$ and afterward the uncertainty further increases. Although this trend continues for both the N2LOnsv and \nsv{N2LL}nsv orders, in general, we observe a slightly better PDF uncertainty for the latter. This is also reflected in the right panel of \fig{fig:rap-pdf} where instead of normalising \eq{eq:rap-pdfuncer} by $\left[\frac{\df \sigma}{\df y}\right]_0 $, we normalize it by the full N2LO results i.e.\ we define 
\begin{align}\label{eq:rap-pdfuncer2}
    \delta\text{PDF}_{\rm r}= \bigg(\left[\frac{\df \sigma}{\df y}\right]_{0} \pm \delta \left[\frac{\df \sigma}{\df y}\right]\bigg)\bigg/ \left[\frac{\df \sigma}{\df y}\right]_{\rm N2LO} \times 100\% \,.
\end{align}
We observe a better PDF uncertainty for the resummed case which is particularly important in the higher rapidity region.

%%%Conclusions%%%%%%%%%%%%%%%%%%%%%%%%%%%%%%%%%
\section{Conclusions}\label{sec:CONCLUSION}
%%%%%%%%%%%%%%%%%%%%%%%%%%%%%%%%%%%%%%%%%%%%%%
In this article, we have studied the effects of soft and next-to-soft threshold logarithms on the total cross-section and the rapidity distribution of Higgs production in bottom quark annihilation. We employed the formalism developed in \cite{AH:2020iki,AH:2020qoa} to calculate the necessary ingredients up to the third order in the strong coupling for both the total cross-section and rapidity distribution. 

We compare the new SV+NSV resummed results for the total cross-section against the fixed order and SV resummed results. The NSV improved resummed results show better perturbative convergence. However, at the N3LO level, the differences are minimized indicating a stable perturbative result for this process. This is further supported by the new fourth-order results for the SV cross-section and SV-resummed calculations, which show corrections of around  $-0.1\%$ and $-0.5\%$ respectively, compared to the N3LO. The seven-point scale uncertainty remains at the $2.5\%$ level, of which the major source is the factorization scale uncertainty. Additionally, we provide partial fourth-order analytical NSV results with a new  $\ln^4 \zb$ coefficient.

For the rapidity distribution, we generally observe a similar behavior as the total cross-section. Here, the NSV resummed results show a better perturbative convergence similar to the SV ones and marginally improve over the SV results in terms of scale uncertainties. Particularly in the third order, a relatively flat correction has been observed for all the rapidity values and the scale uncertainty remains at almost the same level as the fixed order. The new N3LOnsv  gets a flat correction of $-0.25\%$ over the N2LO for the entire rapidity spectrum. On the other hand, the resummed NSV ($\overline{\rm N3LL}$) results get a correction of $-1.3\%$  below $y=2.8$ and it gets up to $-2\%$ corrections at higher rapidity values similar to the Drell-Yan process. Our results for the total cross-section and rapidity distribution of the Higgs production process in bottom annihilation are the most precise QCD results available for this process.

%%%%%%%%%%%%%%%%%%%%%%%%%%%%
\section*{Acknowledgements}
%%%%%%%%%%%%%%%%%%%%%%%%%%%%
We are indebted to V.\ Ravindran for insightful discussions. We further thank C. Williams for providing the NNLO rapidity data. The analytical computation has been performed using the symbolic manipulation system {\sc Form} \cite{Vermaseren:2000nd,Ruijl:2017dtg}. The work of G.D. was supported by the Deutsche Forschungsgemeinschaft (DFG, German Research Foundation) under grant 396021762 - TRR 257 (``Particle Physics Phenomenology after the Higgs Discovery").
\appendix
%%%%%%%%%%%%%%%%%%%%%%%%%%%%%%%%%%%%%%%%%%%%%%%%%%%%%%%%%%%%%%%%
\section{Next-to-soft coefficients: Inclusive production}\label{App:NSV-INCLUSIVE-COEFF}
In this appendix, we collect all the NSV coefficient functions up to N3LO as well as the NSV resummed exponents required up to $\overline{\rm N3LL}$. The NSV coefficient function given in \eq{eq:SV-NSV} can be expanded in strong coupling as,
\begin{align}
\Delta_{\calX, b \bar{b}}^{\rm NSV} 
= 
\sum_{i=1}^{\infty} 
\as^i(\mur)
\Delta_{\calX, b \bar{b}}^{\rm NSV,(i)}\,.
\end{align}
The first three coefficients for the Higgs production cross-section in the bottom annihilation read as (setting $\mur=\muf=m_H$),

\small
\begin{align}
\begin{autobreak}
 
\Delta_{\calX, b \bar{b}}^{\rm NSV,(1)} =
%   \Delta_1^{nsv} =

        % C_F \delta(1-z)   \bigg(
        %   - 4
        %   - 6 L_{fr}
        %   + 8 \zeta_2
        %   \bigg)

         {\color{black} \pmb{L_1}} \bm{ \bigg\{ } C_F   \bigg(
          - 16
          \bigg) \bm{ \bigg\} }

       + {\color{black} \pmb{L_0}} \bm{ \bigg\{ } C_F   \bigg(
           8
          % + 8 L_{fr}
          % - 8 L_{qr}
          \bigg) \bm{ \bigg\} }\,,

       % + {\cal D}_1 C_F   \bigg(
       %     16
       %    \bigg)

       % + {\cal D}_0 C_F   \bigg(
       %    - 8 L_{fr}
       %    + 8 L_{qr}
       %    \bigg)\,,
    
\end{autobreak}   \\
\begin{autobreak}
\Delta_{\calX, b \bar{b}}^{\rm NSV,(2)} =
%   \Delta_2^{nsv} =

       %  C_F n_f \delta(1-z)   \bigg(
       %     \frac{8}{9}
       %    + \frac{2}{3} L_{fr}
       %    - 2 L_{fr}^2
       %    + 8 \zeta_3
       %    - \frac{40}{9} \zeta_2
       %    + \frac{16}{3} \zeta_2 L_{fr}
       %    \bigg)

       % + C_F^2 \delta(1-z)   \bigg(
       %     16
       %    + 21 L_{fr}
       %    + 18 L_{fr}^2
       %    - 60 \zeta_3
       %    - 176 \zeta_3 L_{fr}
       %    + 176 \zeta_3 L_{qr}
       %    - 24 \zeta_2 L_{fr}
       %    - 32 \zeta_2 L_{fr}^2
       %    - 24 \zeta_2 L_{qr}
       %    + 64 \zeta_2 L_{qr} L_{fr}
       %    - 32 \zeta_2 L_{qr}^2
       %    + \frac{8}{5} \zeta_2^2
       %    \bigg)

       % + C_A C_F \delta(1-z)   \bigg(
       %     \frac{166}{9}
       %    - \frac{17}{3} L_{fr}
       %    + 11 L_{fr}^2
       %    - 12 L_{qr}
       %    - 8 \zeta_3
       %    + 24 \zeta_3 L_{fr}
       %    - 24 \zeta_3 L_{qr}
       %    + \frac{232}{9} \zeta_2
       %    - \frac{88}{3} \zeta_2 L_{fr}
       %    - \frac{12}{5} \zeta_2^2
       %    \bigg)

        {\color{black} \pmb{L_3}}  \bm{ \bigg\{ } C_F^2   \bigg(
          - 128
          \bigg) \bm{ \bigg\} }

       + {\color{black} \pmb{L_2}}  \bm{ \bigg\{ } C_F n_f   \bigg(
          - \frac{32}{3}
          \bigg)

       + C_F^2   \bigg(
           248
          % + 192 L_{fr}
          % - 192 L_{qr}
          \bigg)

       +  C_A C_F   \bigg( \frac{176}{3} \bigg) \bm{ \bigg\} }

       + {\color{black} \pmb{L_1}} \bm{ \bigg\{ } C_F n_f   \bigg(
           \frac{352}{9}
          % - \frac{32}{3} L_{qr}
          \bigg)

       + C_F^2   \bigg(
           92
          % - 128 L_{fr}
          % - 64 L_{fr}^2
          % + 224 L_{qr}
          % + 128 L_{qr} L_{fr}
          % - 64 L_{qr}^2
          + 128 \zeta_2
          \bigg)

       + C_A C_F   \bigg(
          - \frac{1948}{9}
        %  + \frac{176}{3} L_{qr}
          + 32 \zeta_2
          \bigg) \bm{ \bigg\} }

       + {\color{black} \pmb{L_0}} \bm{ \bigg\{ }C_F n_f   \bigg(
          - \frac{656}{27}
          % - \frac{128}{9} L_{fr}
          % + \frac{8}{3} L_{fr}^2
          % + \frac{176}{9} L_{qr}
          % - \frac{8}{3} L_{qr}^2
          + \frac{32}{3} \zeta_2
          \bigg)

       + C_F^2   \bigg(
          - 32
          % - 104 L_{fr}
          % - 16 L_{fr}^2
          % + 56 L_{qr}
          % - 16 L_{qr} L_{fr}
          % + 32 L_{qr}^2
          - 256 \zeta_3
          - 128 \zeta_2
          % - 64 \zeta_2 L_{fr}
          % + 64 \zeta_2 L_{qr}
          \bigg)

       +  C_A C_F   \bigg(
           \frac{2804}{27}
          % + \frac{800}{9} L_{fr}
          % - \frac{44}{3} L_{fr}^2
          % - \frac{1064}{9} L_{qr}
          % + \frac{44}{3} L_{qr}^2
          - 56 \zeta_3
          - \frac{224}{3} \zeta_2
          % - 16 \zeta_2 L_{fr}
          % + 16 \zeta_2 L_{qr}
          \bigg) \bm{ \bigg\} } \,,

       % + {\cal D}_3 C_F^2   \bigg(
       %     128
       %    \bigg)

       % + {\cal D}_2 C_F n_f   \bigg(
       %     \frac{32}{3}
       %    \bigg)

       % + {\cal D}_2 C_F^2   \bigg(
       %    - 192 L_{fr}
       %    + 192 L_{qr}
       %    \bigg)

       % + {\cal D}_2 C_A C_F   \bigg(
       %    - \frac{176}{3}
       %    \bigg)

       % + {\cal D}_1 C_F n_f   \bigg(
       %    - \frac{160}{9}
       %    + \frac{32}{3} L_{qr}
       %    \bigg)

       % + {\cal D}_1 C_F^2   \bigg(
       %    - 64
       %    - 96 L_{fr}
       %    + 64 L_{fr}^2
       %    - 128 L_{qr} L_{fr}
       %    + 64 L_{qr}^2
       %    - 128 \zeta_2
       %    \bigg)

       % + {\cal D}_1 C_A C_F   \bigg(
       %     \frac{1072}{9}
       %    - \frac{176}{3} L_{qr}
       %    - 32 \zeta_2
       %    \bigg)

       % + {\cal D}_0 C_F n_f   \bigg(
       %     \frac{224}{27}
       %    + \frac{80}{9} L_{fr}
       %    - \frac{8}{3} L_{fr}^2
       %    - \frac{80}{9} L_{qr}
       %    + \frac{8}{3} L_{qr}^2
       %    - \frac{32}{3} \zeta_2
       %    \bigg)

       % + {\cal D}_0 C_F^2   \bigg(
       %     32 L_{fr}
       %    + 48 L_{fr}^2
       %    - 32 L_{qr}
       %    - 48 L_{qr} L_{fr}
       %    + 256 \zeta_3
       %    + 64 \zeta_2 L_{fr}
       %    - 64 \zeta_2 L_{qr}
       %    \bigg)

       % + {\cal D}_0 C_A C_F   \bigg(
       %    - \frac{1616}{27}
       %    - \frac{536}{9} L_{fr}
       %    + \frac{44}{3} L_{fr}^2
       %    + \frac{536}{9} L_{qr}
       %    - \frac{44}{3} L_{qr}^2
       %    + 56 \zeta_3
       %    + \frac{176}{3} \zeta_2
       %    + 16 \zeta_2 L_{fr}
       %    - 16 \zeta_2 L_{qr}
       %    \bigg)\,,
\end{autobreak}  \\
\begin{autobreak}
\Delta_{\calX, b \bar{b}}^{\rm NSV,(3)} =

         {\color{black} \pmb{L_5}} \bm{ \bigg\{ } C_F^3   \bigg(
          - 512
          \bigg) \bm{ \bigg\} }

       + {\color{black} \pmb{L_4}} \bm{ \bigg\{ }C_F^2 n_f   \bigg(
          - \frac{1280}{9}
          \bigg)

       +  C_F^3   \bigg(
           1728
          % + 1280 L_{fr}
          % - 1280 L_{qr}
          \bigg)

       +  C_A C_F^2   \bigg(
           \frac{7040}{9}
          \bigg) \bm{ \bigg\} }

       + {\color{black} \pmb{L_3}} \bm{ \bigg\{ }C_F n_f^2   \bigg(
          - \frac{256}{27}
          \bigg)

       +  C_F^2 n_f   \bigg(
           \frac{19456}{27}
          % + \frac{1024}{9} L_{fr}
          % - \frac{2560}{9} L_{qr}
          \bigg)

       +  C_F^3   \bigg(
           736
          % - 2496 L_{fr}
          % - 1024 L_{fr}^2
          % + 3264 L_{qr}
          % + 2048 L_{qr} L_{fr}
          % - 1024 L_{qr}^2
          + 3072 \zeta_2
          \bigg)

       + C_A C_F n_f   \bigg(
           \frac{2816}{27}
          \bigg)

       +  C_A C_F^2   \bigg(
          - \frac{111904}{27}
          % - \frac{5632}{9} L_{fr}
          % + \frac{14080}{9} L_{qr}
          + 512 \zeta_2
          \bigg)

       +  C_A^2 C_F   \bigg(
          - \frac{7744}{27}
          \bigg) \bm{ \bigg\} }

       + {\color{black} \pmb{L_2}} \bm{ \bigg\{ } C_F n_f^2   \bigg(
           \frac{1408}{27}
         % - \frac{128}{9} L_{qr}
          \bigg)

       +  C_F^2 n_f   \bigg(
          - \frac{7280}{9}
          + 32 
          % - \frac{1984}{3} L_{fr}
          % + 64 L_{fr}^2
          % + 1056 L_{qr}
          % + 128 L_{qr} L_{fr}
          % - 192 L_{qr}^2
          + \frac{2048}{3} \zeta_2
          \bigg)

       +  C_F^3   \bigg(
          - \frac{2944}{3}
          % - 2672 L_{fr}
          % + 640 L_{fr}^2
          % + 256 L_{fr}^3
          % + 1184 L_{qr}
          % - 2432 L_{qr} L_{fr}
          % - 768 L_{qr} L_{fr}^2
          % + 1792 L_{qr}^2
          % + 768 L_{qr}^2 L_{fr}
          % - 256 L_{qr}^3
          - 10240 \zeta_3
          - 6464 \zeta_2
          % - 4608 \zeta_2 L_{fr}
          % + 4608 \zeta_2 L_{qr}
          \bigg)

       +  C_A C_F n_f   \bigg(
          - \frac{17876}{27}
          + \frac{160}{3} 
        %  + \frac{1408}{9} L_{qr}
          + \frac{128}{3} \zeta_2
          \bigg)

       +  C_A C_F^2   \bigg(
           \frac{41744}{9}
          % + \frac{11968}{3} L_{fr}
          % - 352 L_{fr}^2
          % - 6160 L_{qr}
          % - 704 L_{qr} L_{fr}
          % + 1056 L_{qr}^2
          - 1344 \zeta_3
          - 4744 \zeta_2
          % - 768 \zeta_2 L_{fr}
          % + 768 \zeta_2 L_{qr}
          \bigg)

       +  C_A^2 C_F   \bigg(
           \frac{45628}{27}
        %  - \frac{3872}{9} L_{qr}
          - \frac{724}{3} \zeta_2
          \bigg) \bm{ \bigg\} }

       + {\color{black} \pmb{L_1}} \bm{ \bigg\{ } C_F n_f^2   \bigg( - \frac{6016}{81}  
       % + \frac{1408}{27} L_{qr}
       %    - \frac{64}{9} L_{qr}^2
          + \frac{256}{9} \zeta_2
          \bigg)

       +  C_F^2 n_f   \bigg(
           \frac{27704}{81}
          - 64
          % + \frac{11936}{27} L_{fr}
          % + \frac{1184}{9} L_{fr}^2
          % - \frac{128}{3} L_{fr}^3
          % - \frac{17264}{27} L_{qr}
          % - \frac{3904}{9} L_{qr} L_{fr}
          % + \frac{128}{3} L_{qr} L_{fr}^2
          % + \frac{3584}{9} L_{qr}^2
          % + \frac{128}{3} L_{qr}^2 L_{fr}
          % - \frac{128}{3} L_{qr}^3
          - 1280 \zeta_3
          - \frac{12352}{9} \zeta_2
          % - \frac{1280}{3} \zeta_2 L_{fr}
          % + \frac{1792}{3} \zeta_2 L_{qr}
          \bigg)

       +  C_F^3   \bigg(
          - \frac{1160}{3}
          % + 392 L_{fr}
          % + 1504 L_{fr}^2
          % + 128 L_{fr}^3
          % - 896 L_{qr}
          % - 2240 L_{qr} L_{fr}
          % + 448 L_{qr}^2
          % - 384 L_{qr}^2 L_{fr}
          % + 256 L_{qr}^3
          + 15040 \zeta_3
          % + 11008 \zeta_3 L_{fr}
          % - 11008 \zeta_3 L_{qr}
          - \frac{4160}{3} \zeta_2
          % + 5120 \zeta_2 L_{fr}
          % + 2048 \zeta_2 L_{fr}^2
          % - 5888 \zeta_2 L_{qr}
          % - 4096 \zeta_2 L_{qr} L_{fr}
          % + 2048 \zeta_2 L_{qr}^2
          + \frac{14208}{5} \zeta_2^2
          \bigg)

       +  C_A C_F n_f   \bigg(
           \frac{85712}{81}
          - \frac{320}{3} 
          % - \frac{16976}{27} L_{qr}
          % + \frac{704}{9} L_{qr}^2
          - \frac{4088}{9} \zeta_2
      %    + \frac{128}{3} \zeta_2 L_{qr}
          \bigg)

       +  C_A C_F^2   \bigg(
          - \frac{76976}{81}
          % - \frac{75032}{27} L_{fr}
          % - \frac{8048}{9} L_{fr}^2
          % + \frac{704}{3} L_{fr}^3
          % + \frac{105848}{27} L_{qr}
          % + \frac{24544}{9} L_{qr} L_{fr}
          % - \frac{704}{3} L_{qr} L_{fr}^2
          % - \frac{21248}{9} L_{qr}^2
          % - \frac{704}{3} L_{qr}^2 L_{fr}
          % + \frac{704}{3} L_{qr}^3
          + \frac{21824}{3} \zeta_3
          % + 512 \zeta_3 L_{fr}
          % - 512 \zeta_3 L_{qr}
          + \frac{69064}{9} \zeta_2
          % + \frac{9152}{3} \zeta_2 L_{fr}
          % + 256 \zeta_2 L_{fr}^2
          % - \frac{12544}{3} \zeta_2 L_{qr}
          % - 512 \zeta_2 L_{qr} L_{fr}
          % + 256 \zeta_2 L_{qr}^2
          - \frac{3648}{5} \zeta_2^2
          \bigg)

       +  C_A^2 C_F   \bigg(
          - \frac{235360}{81}
          % + \frac{47752}{27} L_{qr}
          % - \frac{1936}{9} L_{qr}^2
          + 752 \zeta_3
          + \frac{15184}{9} \zeta_2
         % - \frac{704}{3} \zeta_2 L_{qr}
          - \frac{704}{5} \zeta_2^2
          \bigg) \bm{ \bigg\} }

       + {\color{black} \pmb{L_0}} \bm{ \bigg\{ } C_F n_f^2   \bigg(  \frac{25312}{729}  
       % + \frac{128}{27} L_{fr}
       %    - \frac{256}{27} L_{fr}^2
       %    + \frac{32}{27} L_{fr}^3
       %    - \frac{3008}{81} L_{qr}
       %    + \frac{352}{27} L_{qr}^2
       %    - \frac{32}{27} L_{qr}^3
          - \frac{320}{27} \zeta_3
          - \frac{1408}{27} \zeta_2
         % + \frac{128}{9} \zeta_2 L_{qr}
          \bigg)

       +  C_F^2 n_f   \bigg(
          - \frac{5038}{27}
          % + \frac{2396}{27} L_{fr}
          % - \frac{136}{9} L_{fr}^2
          % - \frac{32}{3} L_{fr}^3
          % + \frac{1540}{27} L_{qr}
          % + \frac{176}{9} L_{qr} L_{fr}
          % - \frac{16}{3} L_{qr} L_{fr}^2
          % - \frac{472}{9} L_{qr}^2
          % - \frac{16}{3} L_{qr}^2 L_{fr}
          % + \frac{64}{3} L_{qr}^3
          + \frac{14816}{9} \zeta_3
          % + \frac{896}{3} \zeta_3 L_{fr}
          % - 640 \zeta_3 L_{qr}
          + \frac{2992}{9} \zeta_2
          % + \frac{3520}{9} \zeta_2 L_{fr}
          % + \frac{64}{3} \zeta_2 L_{fr}^2
          % - \frac{5632}{9} \zeta_2 L_{qr}
          % - 128 \zeta_2 L_{qr} L_{fr}
          % + \frac{320}{3} \zeta_2 L_{qr}^2
          + \frac{1472}{15} \zeta_2^2
          \bigg)

       +  C_F^3   \bigg(
           174
          % + 380 L_{fr}
          % + 328 L_{fr}^2
          % - 48 L_{fr}^3
          % - 212 L_{qr}
          % - 56 L_{qr} L_{fr}
          % + 240 L_{qr} L_{fr}^2
          % - 128 L_{qr}^2
          % - 192 L_{qr}^2 L_{fr}
          - 12288 \zeta_5
          + 1088 \zeta_3
          % - 6048 \zeta_3 L_{fr}
          % - 2432 \zeta_3 L_{fr}^2
          % + 7584 \zeta_3 L_{qr}
          % + 4864 \zeta_3 L_{qr} L_{fr}
          % - 2432 \zeta_3 L_{qr}^2
          + \frac{2240}{3} \zeta_2
          % + 1600 \zeta_2 L_{fr}
          % - 704 \zeta_2 L_{fr}^2
          % - 256 \zeta_2 L_{fr}^3
          % - 832 \zeta_2 L_{qr}
          % + 1792 \zeta_2 L_{qr} L_{fr}
          % + 768 \zeta_2 L_{qr} L_{fr}^2
          % - 1088 \zeta_2 L_{qr}^2
          % - 768 \zeta_2 L_{qr}^2 L_{fr}
          % + 256 \zeta_2 L_{qr}^3
          + 6144 \zeta_2 \zeta_3
          - \frac{32576}{15} \zeta_2^2
          % - \frac{7104}{5} \zeta_2^2 L_{fr}
          % + \frac{7104}{5} \zeta_2^2 L_{qr}
          \bigg)

       + C_A C_F n_f   \bigg(
          - \frac{311804}{729}
          % - \frac{1448}{9} L_{fr}
          % + \frac{3368}{27} L_{fr}^2
          % - \frac{352}{27} L_{fr}^3
          % + \frac{40840}{81} L_{qr}
          % - \frac{4424}{27} L_{qr}^2
          % + \frac{352}{27} L_{qr}^3
          + \frac{2584}{9} \zeta_3
          % - \frac{224}{3} \zeta_3 L_{fr}
          + \frac{52624}{81} \zeta_2
          % + \frac{416}{9} \zeta_2 L_{fr}
          % - \frac{32}{3} \zeta_2 L_{fr}^2
          % - 224 \zeta_2 L_{qr}
          % + \frac{32}{3} \zeta_2 L_{qr}^2
          - \frac{736}{15} \zeta_2^2
          \bigg)

       + C_A C_F^2   \bigg(
          - \frac{15790}{27}
          % - \frac{16172}{27} L_{fr}
          % + \frac{1192}{9} L_{fr}^2
          % + \frac{176}{3} L_{fr}^3
          % - \frac{1300}{27} L_{qr}
          % - \frac{2648}{9} L_{qr} L_{fr}
          % + \frac{88}{3} L_{qr} L_{fr}^2
          % + \frac{3832}{9} L_{qr}^2
          % + \frac{88}{3} L_{qr}^2 L_{fr}
          % - \frac{352}{3} L_{qr}^3
          - \frac{22712}{3} \zeta_3
          % - \frac{2480}{3} \zeta_3 L_{fr}
          % + 192 \zeta_3 L_{fr}^2
          % + 3040 \zeta_3 L_{qr}
          % - 384 \zeta_3 L_{qr} L_{fr}
          % + 192 \zeta_3 L_{qr}^2
          - \frac{21472}{9} \zeta_2
          % - \frac{18256}{9} \zeta_2 L_{fr}
          % - \frac{448}{3} \zeta_2 L_{fr}^2
          % + \frac{30736}{9} \zeta_2 L_{qr}
          % + 864 \zeta_2 L_{qr} L_{fr}
          % - \frac{2144}{3} \zeta_2 L_{qr}^2
          + 1472 \zeta_2 \zeta_3
          + \frac{992}{5} \zeta_2^2
          % + \frac{1824}{5} \zeta_2^2 L_{fr}
          % - \frac{1824}{5} \zeta_2^2 L_{qr}
          \bigg)

       +  C_A^2 C_F   \bigg(
           \frac{1431094}{729}
          % + \frac{17164}{27} L_{fr}
          % - \frac{10024}{27} L_{fr}^2
          % + \frac{968}{27} L_{fr}^3
          % - \frac{120524}{81} L_{qr}
          % + \frac{12928}{27} L_{qr}^2
          % - \frac{968}{27} L_{qr}^3
          + 384 \zeta_5
          - \frac{59752}{27} \zeta_3
          % + \frac{176}{3} \zeta_3 L_{fr}
          % + 352 \zeta_3 L_{qr}
          - \frac{165736}{81} \zeta_2
          % - \frac{2672}{9} \zeta_2 L_{fr}
          % + \frac{176}{3} \zeta_2 L_{fr}^2
          % + \frac{7600}{9} \zeta_2 L_{qr}
          % - \frac{176}{3} \zeta_2 L_{qr}^2
          + \frac{352}{3} \zeta_2 \zeta_3
          + 240 \zeta_2^2
          % + \frac{352}{5} \zeta_2^2 L_{fr}
          % - \frac{352}{5} \zeta_2^2 L_{qr}
          \bigg)
          \bm{ \bigg\} } \,.

\end{autobreak}  
\end{align}
Here ${L}_k=\ln^k(\zb)$ and $\zeta_i$ are the Riemann zeta functions. 
% \end{document}

The NSV resummed exponents given in \eq{PsiNSVN}-\eq{hNSV} as required for the ${\rm \overline{N3LL}}$ resummation in $\NB$ scheme read as,
\small
\begin{align}
\begin{autobreak}

   \overline{g}^b_1 =

         \frac{1}{\beta_0} C_F   \bigg(
           4 L_\lambda
          \bigg)\,,
\end{autobreak}    \\
\begin{autobreak}
   \overline{g}^b_2 =

         \frac{1}{(1-\lambda)}\bm{ \bigg\{ }  \frac{1}{\beta_0^2} C_F C_A n_f   \bigg(
          - \frac{40}{3} \lambda
          - \frac{40}{3} L_\lambda
          \bigg)

       + \frac{1}{\beta_0^2} C_F C_A^2   \bigg(
           \frac{136}{3} \lambda
          + \frac{136}{3} L_\lambda
          \bigg)

       + \frac{1}{\beta_0^2} C_F^2 n_f   \bigg( - 8 \lambda
          - 8 L_\lambda \bigg)

       + \frac{1}{\beta_0} C_F n_f   \bigg(
           \frac{40}{9} \lambda
          \bigg)

       + \frac{1}{\beta_0} C_F C_A   \bigg(
          - \frac{268}{9} \lambda
          + 8 \lambda \zeta_2
          \bigg)

       + C_F   \bigg( - 8
%           + 4 L_{qr}
%           - 4 L_{fr}
%           + 4 L_{fr} \lambda
          \bigg) \bm{ \bigg\} } \,,
\end{autobreak}    \\
\begin{autobreak}
   \overline{g}^b_3 =

        \frac{1}{(1-\lambda)^2} \bm{ \bigg\{ }  \frac{1}{\beta_0^3} C_F C_A^2 n_f^2   \bigg(
           \frac{200}{9} \lambda^2
          - \frac{200}{9} L_\lambda^2
          \bigg)

       + \frac{1}{\beta_0^3} C_F C_A^3 n_f   \bigg(
          - \frac{1360}{9} \lambda^2
          + \frac{1360}{9} L_\lambda^2
          \bigg)

       + \frac{1}{\beta_0^3} C_F C_A^4   \bigg(
           \frac{2312}{9} \lambda^2
          - \frac{2312}{9} L_\lambda^2
          \bigg)

       + \frac{1}{\beta_0^3} C_F^2 C_A n_f^2   \bigg(
            \frac{80}{3} \lambda^2
          - \frac{80}{3} L_\lambda^2
          \bigg)

       + \frac{1}{\beta_0^3} C_F^2 C_A^2 n_f   \bigg( - \frac{272}{3} \lambda^2
          + \frac{272}{3} L_\lambda^2 \bigg)

       + \frac{1}{\beta_0^3} C_F^3 n_f^2   \bigg(
            8 \lambda^2
          - 8 L_\lambda^2
          \bigg)

       + \frac{1}{\beta_0^2} C_F C_A n_f^2   \bigg(
            \frac{400}{27} \lambda
          - \frac{31}{3} \lambda^2
          + \frac{400}{27} L_\lambda
          \bigg)

       + \frac{1}{\beta_0^2} C_F C_A^2 n_f   \bigg(
          - \frac{4040}{27} \lambda
          + \frac{80}{3} \lambda \zeta_2
          + \frac{1145}{9} \lambda^2
          - \frac{40}{3} \lambda^2 \zeta_2
          - \frac{4040}{27} L_\lambda
          + \frac{80}{3} L_\lambda \zeta_2
          \bigg)

       + \frac{1}{\beta_0^2} C_F C_A^3   \bigg(
            \frac{9112}{27} \lambda
          - \frac{272}{3} \lambda \zeta_2
          - \frac{2471}{9} \lambda^2
          + \frac{136}{3} \lambda^2 \zeta_2
          + \frac{9112}{27} L_\lambda
          - \frac{272}{3} L_\lambda \zeta_2
          \bigg)

       + \frac{1}{\beta_0^2} C_F^2 n_f^2   \bigg(
           \frac{80}{9} \lambda
          - \frac{62}{9} \lambda^2
          + \frac{80}{9} L_\lambda
          \bigg)

       + \frac{1}{\beta_0^2} C_F^2 C_A n_f   \bigg(
          - \frac{536}{9} \lambda
          + 16 \lambda \zeta_2
          + \frac{473}{9} \lambda^2
          - 8 \lambda^2 \zeta_2
          - \frac{536}{9} L_\lambda
          + 16 L_\lambda \zeta_2
          \bigg)

       + \frac{1}{\beta_0^2} C_F^3 n_f   \bigg(
          - 2 \lambda^2
          \bigg)

       + \frac{1}{\beta_0} C_F n_f^2   \bigg( \frac{16}{27} \lambda
          - \frac{8}{27} \lambda^2  \bigg)

       + \frac{1}{\beta_0} C_F C_A n_f   \bigg(
           \frac{836}{27} \lambda
          + \frac{112}{3} \lambda \zeta_3
          - \frac{160}{9} \lambda \zeta_2
          - \frac{418}{27} \lambda^2
          - \frac{56}{3} \lambda^2 \zeta_3
          + \frac{80}{9} \lambda^2 \zeta_2
          - \frac{80}{3} L_\lambda
%           + \frac{40}{3} L_\lambda L_{qr}
          \bigg)

       + \frac{1}{\beta_0} C_F C_A^2   \bigg(
          - \frac{490}{3} \lambda
          - \frac{88}{3} \lambda \zeta_3
          + \frac{1072}{9} \lambda \zeta_2
          - \frac{176}{5} \lambda \zeta_2^2
          + \frac{245}{3} \lambda^2
          + \frac{44}{3} \lambda^2 \zeta_3
          - \frac{536}{9} \lambda^2 \zeta_2
          + \frac{88}{5} \lambda^2 \zeta_2^2
          + \frac{272}{3} L_\lambda
%           - \frac{136}{3} L_\lambda L_{qr}
          \bigg)

       + \frac{1}{\beta_0} C_F^2 n_f   \bigg(
           \frac{110}{3} \lambda
          - 32 \lambda \zeta_3
          - \frac{55}{3} \lambda^2
          + 16 \lambda^2 \zeta_3
          - 16 L_\lambda
%           + 8 L_\lambda L_{qr}
          \bigg)

       + C_F n_f   \bigg(
           \frac{352}{27}
%           - \frac{88}{9} L_{qr}
%           + \frac{4}{3} L_{qr}^2
%           + \frac{40}{9} L_{fr}
%           - \frac{80}{9} L_{fr} \lambda
%           + \frac{40}{9} L_{fr} \lambda^2
%           - \frac{4}{3} L_{fr}^2
%           + \frac{8}{3} L_{fr}^2 \lambda
%           - \frac{4}{3} L_{fr}^2 \lambda^2
    
          \bigg)

       + C_F C_A   \bigg(- \frac{2416}{27}
          + 28 \zeta_3
          + 16 \zeta_2
%           + \frac{532}{9} L_{qr}
%           - 8 L_{qr} \zeta_2
%           - \frac{22}{3} L_{qr}^2
%           - \frac{268}{9} L_{fr}
%           + 8 L_{fr} \zeta_2
%           + \frac{536}{9} L_{fr} \lambda
%           - 16 L_{fr} \lambda \zeta_2
%           - \frac{268}{9} L_{fr} \lambda^2
%           + 8 L_{fr} \lambda^2 \zeta_2
%           + \frac{22}{3} L_{fr}^2
%           - \frac{44}{3} L_{fr}^2 \lambda
%           + \frac{22}{3} L_{fr}^2 \lambda^2
          \bigg) \bm{ \bigg\} }  \,,
\end{autobreak} \\
\begin{autobreak}
 
  \overline{g}^b_4 = 

   \frac{1}{6\beta_0^4 (\lambda-1)^3} \bm{ \bigg\{ }  \beta_0^2 \beta_2 C_F n_f   \bigg(
          - \frac{80}{3} \lambda^2
          + \frac{80}{9} \lambda^3
          \bigg)

       + \beta_0^2 \beta_2 C_F C_A   \bigg(
           \frac{536}{3} \lambda^2
          - 48 \lambda^2 \zeta_2
          - \frac{536}{9} \lambda^3
          + 16 \lambda^3 \zeta_2
          \bigg)

       + \beta_0^2 \beta_3 C_F   \bigg(
           12 \lambda^2
          - 8 \lambda^3
          \bigg)

       + \beta_0^3 n_f \frac{d_F^{abcd} d_F^{abcd}}{N_F}   \bigg(
          - 2560 \lambda \zeta_5
          - 512 \lambda \zeta_3
          + 1536 \lambda \zeta_2
          + 2560 \lambda^2 \zeta_5
          + 512 \lambda^2 \zeta_3
          - 1536 \lambda^2 \zeta_2
          - \frac{2560}{3} \lambda^3 \zeta_5
          - \frac{512}{3} \lambda^3 \zeta_3
          + 512 \lambda^3 \zeta_2
          \bigg)

       + \beta_0^3 \frac{d_F^{abcd} d_A^{abcd}}{N_F}   \bigg(
           7040 \lambda \zeta_5
          + 256 \lambda \zeta_3
          - 2304 \lambda \zeta_3^2
          - 768 \lambda \zeta_2
          - \frac{47616}{35} \lambda \zeta_2^3
          - 7040 \lambda^2 \zeta_5
          - 256 \lambda^2 \zeta_3
          + 2304 \lambda^2 \zeta_3^2
          + 768 \lambda^2 \zeta_2
          + \frac{47616}{35} \lambda^2 \zeta_2^3
          + \frac{7040}{3} \lambda^3 \zeta_5
          + \frac{256}{3} \lambda^3 \zeta_3
          - 768 \lambda^3 \zeta_3^2
          - 256 \lambda^3 \zeta_2
          - \frac{15872}{35} \lambda^3 \zeta_2^3
          \bigg)

       + \beta_0^3 C_F n_f^3   \bigg(
          - \frac{64}{27} \lambda
          + \frac{128}{9} \lambda \zeta_3
          + \frac{64}{27} \lambda^2
          - \frac{128}{9} \lambda^2 \zeta_3
          - \frac{64}{81} \lambda^3
          + \frac{128}{27} \lambda^3 \zeta_3
          \bigg)

       + \beta_0^3 C_F C_A n_f^2   \bigg(
           \frac{1846}{27} \lambda
          + \frac{4480}{9} \lambda \zeta_3
          - \frac{1216}{27} \lambda \zeta_2
          - \frac{448}{5} \lambda \zeta_2^2
          - \frac{1846}{27} \lambda^2
          - \frac{4480}{9} \lambda^2 \zeta_3
          + \frac{1216}{27} \lambda^2 \zeta_2
          + \frac{448}{5} \lambda^2 \zeta_2^2
          + \frac{1846}{81} \lambda^3
          + \frac{4480}{27} \lambda^3 \zeta_3
          - \frac{1216}{81} \lambda^3 \zeta_2
          - \frac{448}{15} \lambda^3 \zeta_2^2
          \bigg)

       + \beta_0^3 C_F C_A^2 n_f   \bigg(
          - \frac{48274}{27} \lambda
          + \frac{4192}{3} \lambda \zeta_5
          - \frac{46208}{9} \lambda \zeta_3
          + \frac{40640}{27} \lambda \zeta_2
          + 896 \lambda \zeta_2 \zeta_3
          - \frac{704}{5} \lambda \zeta_2^2
          + \frac{48274}{27} \lambda^2
          - \frac{4192}{3} \lambda^2 \zeta_5
          + \frac{46208}{9} \lambda^2 \zeta_3
          - \frac{40640}{27} \lambda^2 \zeta_2
          - 896 \lambda^2 \zeta_2 \zeta_3
          + \frac{704}{5} \lambda^2 \zeta_2^2
          - \frac{48274}{81} \lambda^3
          + \frac{4192}{9} \lambda^3 \zeta_5
          - \frac{46208}{27} \lambda^3 \zeta_3
          + \frac{40640}{81} \lambda^3 \zeta_2
          + \frac{896}{3} \lambda^3 \zeta_2 \zeta_3
          - \frac{704}{15} \lambda^3 \zeta_2^2
          \bigg)

       + \beta_0^3 C_F C_A^3   \bigg(
           \frac{168556}{27} \lambda
          - \frac{7216}{3} \lambda \zeta_5
          + \frac{41888}{9} \lambda \zeta_3
          - 96 \lambda \zeta_3^2
          - \frac{176800}{27} \lambda \zeta_2
          - 704 \lambda \zeta_2 \zeta_3
          + \frac{21648}{5} \lambda \zeta_2^2
          - \frac{40064}{35} \lambda \zeta_2^3
          - \frac{168556}{27} \lambda^2
          + \frac{7216}{3} \lambda^2 \zeta_5
          - \frac{41888}{9} \lambda^2 \zeta_3
          + 96 \lambda^2 \zeta_3^2
          + \frac{176800}{27} \lambda^2 \zeta_2
          + 704 \lambda^2 \zeta_2 \zeta_3
          - \frac{21648}{5} \lambda^2 \zeta_2^2
          + \frac{40064}{35} \lambda^2 \zeta_2^3
          + \frac{168556}{81} \lambda^3
          - \frac{7216}{9} \lambda^3 \zeta_5
          + \frac{41888}{27} \lambda^3 \zeta_3
          - 32 \lambda^3 \zeta_3^2
          - \frac{176800}{81} \lambda^3 \zeta_2
          - \frac{704}{3} \lambda^3 \zeta_2 \zeta_3
          + \frac{7216}{5} \lambda^3 \zeta_2^2
          - \frac{40064}{105} \lambda^3 \zeta_2^3
          \bigg)

       + \beta_0^3 C_F^2 n_f^2   \bigg(
           \frac{4784}{27} \lambda
          - \frac{1280}{3} \lambda \zeta_3
          + \frac{384}{5} \lambda \zeta_2^2
          - \frac{4784}{27} \lambda^2
          + \frac{1280}{3} \lambda^2 \zeta_3
          - \frac{384}{5} \lambda^2 \zeta_2^2
          + \frac{4784}{81} \lambda^3
          - \frac{1280}{9} \lambda^3 \zeta_3
          + \frac{128}{5} \lambda^3 \zeta_2^2
          \bigg)

       + \beta_0^3 C_F^2 C_A n_f   \bigg(
          - \frac{68132}{27} \lambda
          + 960 \lambda \zeta_5
          + \frac{7424}{3} \lambda \zeta_3
          + 880 \lambda \zeta_2
          - 768 \lambda \zeta_2 \zeta_3
          - \frac{2112}{5} \lambda \zeta_2^2
          + \frac{68132}{27} \lambda^2
          - 960 \lambda^2 \zeta_5
          - \frac{7424}{3} \lambda^2 \zeta_3
          - 880 \lambda^2 \zeta_2
          + 768 \lambda^2 \zeta_2 \zeta_3
          + \frac{2112}{5} \lambda^2 \zeta_2^2
          - \frac{68132}{81} \lambda^3
          + 320 \lambda^3 \zeta_5
          + \frac{7424}{9} \lambda^3 \zeta_3
          + \frac{880}{3} \lambda^3 \zeta_2
          - 256 \lambda^3 \zeta_2 \zeta_3
          - \frac{704}{5} \lambda^3 \zeta_2^2
          \bigg)

       + \beta_0^3 C_F^3 n_f   \bigg(
           \frac{1144}{3} \lambda
          - 1920 \lambda \zeta_5
          + 1184 \lambda \zeta_3
          - \frac{1144}{3} \lambda^2
          + 1920 \lambda^2 \zeta_5
          - 1184 \lambda^2 \zeta_3
          + \frac{1144}{9} \lambda^3
          - 640 \lambda^3 \zeta_5
          + \frac{1184}{3} \lambda^3 \zeta_3
          \bigg)

       + \beta_0^3 \beta_2 C_F   \bigg(
           48 \lambda
%           - 24 \lambda L_{qr}
          \bigg)

       + \beta_0^4 C_F n_f^2   \bigg(
          - \frac{5888}{243}
          + \frac{224}{9} \zeta_3
          - \frac{80}{9} \zeta_2
%           - \frac{32}{9} L_{fr}
%           + \frac{32}{9} L_{qr}
%           + \frac{32}{3} \lambda L_{fr}
%           - \frac{32}{3} \lambda^2 L_{fr}
%           + \frac{32}{9} \lambda^3 L_{fr}
          \bigg)

       + \beta_0^4 C_F C_A n_f   \bigg(
          - \frac{113972}{243}
          - \frac{2576}{9} \zeta_3
          + \frac{11416}{27} \zeta_2
          - \frac{576}{5} \zeta_2^2
%           - \frac{1672}{9} L_{fr}
%           - 224 L_{fr} \zeta_3
%           + \frac{320}{3} L_{fr} \zeta_2
%           + \frac{1672}{9} L_{qr}
%           + 224 L_{qr} \zeta_3
%           - \frac{320}{3} L_{qr} \zeta_2
%           + \frac{1672}{3} \lambda L_{fr}
%           + 672 \lambda L_{fr} \zeta_3
%           - 320 \lambda L_{fr} \zeta_2
%           - \frac{1672}{3} \lambda^2 L_{fr}
%           - 672 \lambda^2 L_{fr} \zeta_3
%           + 320 \lambda^2 L_{fr} \zeta_2
%           + \frac{1672}{9} \lambda^3 L_{fr}
%           + 224 \lambda^3 L_{fr} \zeta_3
%           - \frac{320}{3} \lambda^3 L_{fr} \zeta_2
          \bigg)

       + \beta_0^4 C_F C_A^2   \bigg(
           \frac{749842}{243}
          + 1152 \zeta_5
          - 2280 \zeta_3
          - \frac{63892}{27} \zeta_2
          + 352 \zeta_2 \zeta_3
          + \frac{4224}{5} \zeta_2^2
%           + 980 L_{fr}
%           + 176 L_{fr} \zeta_3
%           - \frac{2144}{3} L_{fr} \zeta_2
%           + \frac{1056}{5} L_{fr} \zeta_2^2
%           - 980 L_{qr}
%           - 176 L_{qr} \zeta_3
%           + \frac{2144}{3} L_{qr} \zeta_2
%           - \frac{1056}{5} L_{qr} \zeta_2^2
%           - 2940 \lambda L_{fr}
%           - 528 \lambda L_{fr} \zeta_3
%           + 2144 \lambda L_{fr} \zeta_2
%           - \frac{3168}{5} \lambda L_{fr} \zeta_2^2
%           + 2940 \lambda^2 L_{fr}
%           + 528 \lambda^2 L_{fr} \zeta_3
%           - 2144 \lambda^2 L_{fr} \zeta_2
%           + \frac{3168}{5} \lambda^2 L_{fr} \zeta_2^2
%           - 980 \lambda^3 L_{fr}
%           - 176 \lambda^3 L_{fr} \zeta_3
%           + \frac{2144}{3} \lambda^3 L_{fr} \zeta_2
%           - \frac{1056}{5} \lambda^3 L_{fr} \zeta_2^2
          \bigg)

       + \beta_0^4 C_F^2 n_f   \bigg(
          - \frac{7382}{9}
          + \frac{1760}{3} \zeta_3
          + 24 \zeta_2
          + \frac{192}{5} \zeta_2^2
%           - 220 L_{fr}
%           + 192 L_{fr} \zeta_3
%           + 220 L_{qr}
%           - 192 L_{qr} \zeta_3
%           + 660 \lambda L_{fr}
%           - 576 \lambda L_{fr} \zeta_3
%           - 660 \lambda^2 L_{fr}
%           + 576 \lambda^2 L_{fr} \zeta_3
%           + 220 \lambda^3 L_{fr}
%           - 192 \lambda^3 L_{fr} \zeta_3
          \bigg)

       + \beta_0^5 C_F n_f   \bigg(
          - \frac{4000}{27}
          + 128 \zeta_3
          + \frac{56}{3} \zeta_2
%           + \frac{80}{3} L_{fr}^2
%           + \frac{1408}{9} L_{qr}
%           - 16 L_{qr} \zeta_2
%           - \frac{80}{3} L_{qr}^2
%           - 80 \lambda L_{fr}^2
%           + 80 \lambda^2 L_{fr}^2
%           - \frac{80}{3} \lambda^3 L_{fr}^2
          \bigg)

       + \beta_0^5 C_F C_A   \bigg(
           \frac{29104}{27}
          - 1376 \zeta_3
          - \frac{260}{3} \zeta_2
          - 144 \zeta_2^2
%           - \frac{536}{3} L_{fr}^2
%           + 48 L_{fr}^2 \zeta_2
%           - \frac{9664}{9} L_{qr}
%           + 336 L_{qr} \zeta_3
%           + 280 L_{qr} \zeta_2
%           + \frac{536}{3} L_{qr}^2
%           - 48 L_{qr}^2 \zeta_2
%           + 536 \lambda L_{fr}^2
%           - 144 \lambda L_{fr}^2 \zeta_2
%           - 536 \lambda^2 L_{fr}^2
%           + 144 \lambda^2 L_{fr}^2 \zeta_2
%           + \frac{536}{3} \lambda^3 L_{fr}^2
%           - 48 \lambda^3 L_{fr}^2 \zeta_2
          \bigg)

       + \beta_0^6 C_F   \bigg(
           184 \zeta_3
          + 48 \zeta_2
%           + 8 L_{fr}^3
%           - 24 L_{qr} \zeta_2
%           + 48 L_{qr}^2
%           - 8 L_{qr}^3
%           - 24 \lambda L_{fr}^3
%           + 24 \lambda^2 L_{fr}^3
%           - 8 \lambda^3 L_{fr}^3
          \bigg)

       + \beta_1 \beta_0 \beta_2 C_F   \bigg(
          - 24 \lambda^2
          + 16 \lambda^3
          - 24 L_\lambda \lambda
          \bigg)

       + \beta_1 \beta_0^2 C_F n_f^2   \bigg(
           \frac{32}{9} \lambda
          - \frac{32}{9} \lambda^2
          + \frac{32}{27} \lambda^3
          + \frac{32}{9} L_\lambda
          \bigg)

       + \beta_1 \beta_0^2 C_F C_A n_f   \bigg(
           \frac{1672}{9} \lambda
          + 224 \lambda \zeta_3
          - \frac{320}{3} \lambda \zeta_2
          - \frac{1672}{9} \lambda^2
          - 224 \lambda^2 \zeta_3
          + \frac{320}{3} \lambda^2 \zeta_2
          + \frac{1672}{27} \lambda^3
          + \frac{224}{3} \lambda^3 \zeta_3
          - \frac{320}{9} \lambda^3 \zeta_2
          + \frac{1672}{9} L_\lambda
          + 224 L_\lambda \zeta_3
          - \frac{320}{3} L_\lambda \zeta_2
          \bigg)

       + \beta_1 \beta_0^2 C_F C_A^2   \bigg(
          - 980 \lambda
          - 176 \lambda \zeta_3
          + \frac{2144}{3} \lambda \zeta_2
          - \frac{1056}{5} \lambda \zeta_2^2
          + 980 \lambda^2
          + 176 \lambda^2 \zeta_3
          - \frac{2144}{3} \lambda^2 \zeta_2
          + \frac{1056}{5} \lambda^2 \zeta_2^2
          - \frac{980}{3} \lambda^3
          - \frac{176}{3} \lambda^3 \zeta_3
          + \frac{2144}{9} \lambda^3 \zeta_2
          - \frac{352}{5} \lambda^3 \zeta_2^2
          - 980 L_\lambda
          - 176 L_\lambda \zeta_3
          + \frac{2144}{3} L_\lambda \zeta_2
          - \frac{1056}{5} L_\lambda \zeta_2^2
          \bigg)

       + \beta_1 \beta_0^2 C_F^2 n_f   \bigg(
           220 \lambda
          - 192 \lambda \zeta_3
          - 220 \lambda^2
          + 192 \lambda^2 \zeta_3
          + \frac{220}{3} \lambda^3
          - 64 \lambda^3 \zeta_3
          + 220 L_\lambda
          - 192 L_\lambda \zeta_3
          \bigg)

       + \beta_1 \beta_0^3 C_F n_f   \bigg(
          \frac{1408}{9} L_\lambda
          - 16 L_\lambda \zeta_2
%           - \frac{160}{3} L_\lambda L_{qr}
          \bigg)

       + \beta_1 \beta_0^3 C_F C_A   \bigg(
          - \frac{9664}{9} L_\lambda
          + 336 L_\lambda \zeta_3
          + 280 L_\lambda \zeta_2
%           + \frac{1072}{3} L_\lambda L_{qr}
%           - 96 L_\lambda L_{qr} \zeta_2
          \bigg)

       + \beta_1 \beta_0^4 C_F   \bigg(
           12 \zeta_2
%           - 12 L_{fr}^2
%           - 48 L_{qr}
%           + 12 L_{qr}^2
%           + 36 \lambda L_{fr}^2
%           - 36 \lambda^2 L_{fr}^2
%           + 12 \lambda^3 L_{fr}^2
          - 24 L_\lambda \zeta_2
%           + 96 L_\lambda L_{qr}
%           - 24 L_\lambda L_{qr}^2
          \bigg)

       + \beta_1^2 \beta_0 C_F n_f   \bigg(
           \frac{80}{3} \lambda^2
          - \frac{80}{9} \lambda^3
          - \frac{80}{3} L_\lambda^2
          \bigg)

       + \beta_1^2 \beta_0 C_F C_A   \bigg(
          - \frac{536}{3} \lambda^2
          + 48 \lambda^2 \zeta_2
          + \frac{536}{9} \lambda^3
          - 16 \lambda^3 \zeta_2
          + \frac{536}{3} L_\lambda^2
          - 48 L_\lambda^2 \zeta_2
          \bigg)

       + \beta_1^2 \beta_0^2 C_F   \bigg(
          - 48 \lambda
%           + 24 \lambda L_{qr}
          - 48 L_\lambda
%           + 24 L_\lambda L_{qr}
          + 48 L_\lambda^2
%           - 24 L_\lambda^2 L_{qr}
          \bigg)

       + \beta_1^3 C_F   \bigg(
           12 \lambda^2
          - 8 \lambda^3
          + 24 L_\lambda \lambda
          + 12 L_\lambda^2
          - 8 L_\lambda^3
          \bigg)\bm{ \bigg\} } \,,
\end{autobreak} 
\end{align}

%-------------------------------------------------------------------------------------------------------------------------------------------------
\small

Here, $L_{\lambda} = \ln(1-\lambda)$ with $\lambda = 2  a_s(\mu_R^2) \beta_0 \ln(\overbar{N})$.
%%%%%%%%%%%%%%%%%%%%%%%%%%%%%%%%%%%%%%%%%%%%%%%%%%%%%%%%%
\section{Next-to-soft coefficients: Rapidity}\label{App:NSV-RAPIDITY-COEFF}
%%%%%%%%%%%%%%%%%%%%%%%%%%%%%%%%%%%%%%%%%%%%%%%%%%%%%%%%
The NSV coefficient function for the Higgs rapidity distribution in \eq{eq:SV-NSV} can be expanded in strong coupling as,
\begin{align}
\Delta_{d, b \bar{b}}^{\rm NSV} 
= 
\sum_{i=1}^{\infty} 
\as^i(\mur)
\Delta_{d, b \bar{b}}^{\rm NSV,(i)}\,,
\end{align}%
where
\small
\begin{align}
\begin{autobreak}
%    \Delta_{d,1}^{nsv} =
    \Delta_{d, b \bar{b}}^{\rm NSV,(1)} =

       %  C_F \overline{{\cal D}}_0 {\cal D}_0   \bigg( 
       %     2
       %    \bigg)

       % + C_F \delta(1-z_1) \overline{{\cal D}}_1   \bigg( 
       %     4
       %    \bigg)

       % + C_F \delta(1-z_1) \overline{\delta}   \bigg( 
       %    - 2
       %    + 6 \zeta_2
       %    \bigg) + 
          
          {\color{black} \pmb{\overline{{\cal D}}_0 L_0}} \bm{ \bigg\{ }  C_F   \bigg(
          - 4
          \bigg) \bm{ \bigg\} } 

       + {\color{black} \pmb{\overline{\delta} L_1 }}  \bm{ \bigg\{ }  C_F \bigg(
          - 4
          \bigg) \bm{ \bigg\} } 

       + {\color{black} \pmb{ \overline{\delta} L_0  }} \bm{ \bigg\{ }  C_F \bigg(
          2 
          \bigg) \bm{ \bigg\} } 
          + (z_1 \leftrightarrow z_2) \,,
\end{autobreak}\\
\begin{autobreak}
    
%  \Delta_{d,2}^{nsv} =
 \Delta_{d, b \bar{b}}^{\rm NSV,(2)} =

        {\color{black} \pmb{\overline{{\cal D}}_1  L_0 }} \bm{ \bigg\{ }  C_F n_f \bigg(  - \frac{8}{3} \bigg) + C_F^2  \bigg(   24 \bigg) +  C_A C_F \bigg( 
           \frac{44}{3}
          \bigg) \bm{ \bigg\} }

       +  {\color{black} \pmb{\overline{{\cal D}}_0 L_1 }}  \bm{ \bigg\{ }  C_F n_f \bigg(  - \frac{8}{3}
          \bigg)  + C_F^2   \bigg(  40 \bigg) 
          +  C_A C_F \bigg(  \frac{44}{3} \bigg) \bm{ \bigg\} }

       +  {\color{black} \pmb{\overline{{\cal D}}_0 L_0 }} \bm{ \bigg\{ }   C_F n_f\bigg( 
           \frac{76}{9}
          \bigg) + C_F^2  \bigg( 
           28
          - 16 \zeta_2
          \bigg) +  C_A C_F \bigg( 
          - \frac{466}{9}
          + 8 \zeta_2
          \bigg) \bm{ \bigg\} }

       + {\color{black} \pmb{ \overline{{\cal D}}_1 L_1 }}  \bm{ \bigg\{ }  C_F^2  \bigg( 
          - 48
          \bigg) \bm{ \bigg\} } 

       +  {\color{black} \pmb{\overline{{\cal D}}_2 L_0 }} \bm{ \bigg\{ }   C_F^2 \bigg( 
          - 24
          \bigg) \bm{ \bigg\} } 
 +  {\color{black} \pmb{\overline{{\cal D}}_0 L_2 }} \bm{ \bigg\{ }  C_F^2  \bigg( 
          - 24
          \bigg) \bm{ \bigg\} }

       + {\color{black} \pmb{\overline{\delta} L_3 }} \bm{ \bigg\{ }  C_F^2  \bigg( 
          - 8
          \bigg) \bm{ \bigg\} } 
    +  {\color{black} \pmb{\overline{\delta} L_2 }}  \bm{ \bigg\{ }  C_F n_f\bigg( 
          - \frac{4}{3}
          \bigg) + C_F^2 \bigg( 
           24
          \bigg) + C_A C_F \bigg( 
           \frac{22}{3}
          \bigg) \bm{ \bigg\} }

       + {\color{black} \pmb{\overline{\delta} L_1  }} \bm{ \bigg\{ }   C_F n_f \bigg( 
           \frac{76}{9}
          \bigg) +  C_F^2  \bigg( 
           18
          - 16 \zeta_2
          \bigg) +  C_A C_F \bigg( 
          - \frac{376}{9}
          + 8 \zeta_2
          \bigg)\bm{ \bigg\} }

     +  {\color{black} \pmb{\overline{\delta} L_0  }} \bm{ \bigg\{ } C_F n_f  \bigg( 
          - \frac{268}{27}
          + \frac{8}{3} \zeta_2
          \bigg) +  C_F^2  \bigg( 
          - 8
          - 32 \zeta_3
          - 8 \zeta_2
          \bigg)  +   C_A C_F \bigg( 
           \frac{1000}{27}
          - 28 \zeta_3
          - \frac{56}{3} \zeta_2
          \bigg) \bm{ \bigg\} } 
      +    (z_1 \leftrightarrow z_2) \,,
\end{autobreak}\\
\begin{autobreak}
    
% \Delta_{d,3}^{nsv} =
 \Delta_{d, b \bar{b}}^{\rm NSV,(3)} =

         {\color{black} \pmb{ \overline{{\cal D}}_1 L_1}} \bm{ \bigg\{ } C_F n_f^2  \bigg( 
          - \frac{32}{9}
          \bigg)
   + C_F^3   \bigg( 
         296
          + 192 \zeta_2
          \bigg)

       + C_A C_F n_f   \bigg( 
           \frac{352}{9}
          \bigg) \bm{ \bigg\} } 

       +    {\color{black} \pmb{\overline{{\cal D}}_2  L_0 }} \bm{ \bigg\{ } C_F n_f^2\bigg( 
          - \frac{16}{9}
          \bigg)
  + C_A C_F n_f   \bigg( 
           \frac{176}{9}
          \bigg)
   + C_A C_F^2   \bigg( 
          - 592
          + 96 \zeta_2
          \bigg)\bm{ \bigg\} }

       +  {\color{black} \pmb{\overline{{\cal D}}_1 L_0 }} \bm{ \bigg\{ } C_F n_f^2 \bigg( 
           \frac{304}{27}
          \bigg)
   + C_F^3   \bigg( 
          - 96
          - 640 \zeta_3
          - 288 \zeta_2
          \bigg)
   + C_A C_F^2    \bigg( 
           \frac{1616}{3}
          - 336 \zeta_3
          - 448 \zeta_2
          \bigg)\bm{ \bigg\} } 
          
       + {\color{black} \pmb{\overline{{\cal D}}_0 L_2 }}  \bm{ \bigg\{ } C_F n_f^2\bigg( 
          - \frac{16}{9}
          \bigg)
 + C_F^2 n_f  \bigg( 
           112
          \bigg)
  
       + C_F^3   \bigg( 
          128
          + 96 \zeta_2
          \bigg)
     + C_A C_F n_f  \bigg( 
           \frac{176}{9}
          \bigg)
 + C_A C_F^2   \bigg( 
          - 640
          + 96 \zeta_2
          \bigg)
    + C_A^2 C_F   \bigg( 
          - \frac{484}{9}
          \bigg)\bm{ \bigg\} }

       + {\color{black} \pmb{\overline{{\cal D}}_0 L_1 }}  \bm{ \bigg\{ } C_F n_f^2\bigg( 
           \frac{304}{27}
          \bigg)

       + C_F^2 n_f  \bigg( 
          - \frac{1328}{9}
          + 64 \zeta_2
          \bigg)
   + C_F^3    \bigg( 
          - 160
          - 640 \zeta_3
          - 192 \zeta_2
          \bigg)

       + C_A C_F n_f    \bigg( 
          - \frac{3536}{27}
          + \frac{32}{3} \zeta_2
          \bigg)

       + C_A C_F^2   \bigg( 
          \frac{7652}{9}
          - 336 \zeta_3
          - 512 \zeta_2
          \bigg)

       + C_A^2 C_F   \bigg( 
          \frac{9496}{27}
          - \frac{176}{3} \zeta_2
          \bigg)\bm{ \bigg\} }

       + {\color{black} \pmb{\overline{{\cal D}}_0 L_0 }}  \bm{ \bigg\{ } C_F n_f^2\bigg( 
          - \frac{1264}{81}
          + \frac{32}{9} \zeta_2
          \bigg)

       + C_F^2 n_f   \bigg( 
           \frac{194}{9}
          - \frac{320}{3} \zeta_3
          - \frac{320}{9} \zeta_2
          \bigg)

       + C_F^3   \bigg( 
          - 106
          + 816 \zeta_3
          - 24 \zeta_2
          + \frac{192}{5} \zeta_2^2
          \bigg)

       + C_A C_F n_f    \bigg( 
           \frac{16952}{81}
          - \frac{608}{9} \zeta_2
          \bigg)

       + C_A C_F^2    \bigg( 
           \frac{490}{9}
          + \frac{1160}{3} \zeta_3
          + \frac{848}{9} \zeta_2
          + \frac{48}{5} \zeta_2^2
          \bigg)

       + C_A^2 C_F   \bigg( 
          - \frac{49582}{81}
          + 176 \zeta_3
          + \frac{856}{3} \zeta_2
          - \frac{176}{5} \zeta_2^2
          \bigg)\bm{ \bigg\} }

       +  {\color{black} \pmb{\overline{\delta} L_3 }}  \bm{ \bigg\{ } C_F n_f^2\bigg( 
          - \frac{16}{27}
          \bigg)

       + C_F^2 n_f    \bigg( 
           \frac{1040}{27}
          \bigg)

       + C_F^3    \bigg( 
           36
          + 32 \zeta_2
          \bigg)

       + C_A C_F n_f   \bigg( 
           \frac{176}{27}
          \bigg)

       + C_A C_F^2   \bigg( 
          - \frac{5756}{27}
          + 32 \zeta_2
          \bigg)

       + C_A^2 C_F   \bigg( 
          - \frac{484}{27}
          \bigg)\bm{ \bigg\} }

       +  {\color{black} \pmb{\overline{\delta} L_2  }} \bm{ \bigg\{ } C_F n_f^2\bigg( 
           \frac{152}{27}
          \bigg)

       + C_F^2 n_f   \bigg( 
          - \frac{908}{9}
          + 16
          + 32 \zeta_2
          \bigg)

   + C_F^3    \bigg( 
          - \frac{272}{3}
          - 320 \zeta_3
          - 96 \zeta_2
          \bigg)

       + C_A C_F n_f    \bigg( 
          - \frac{2398}{27}
          + \frac{80}{3} 
          + \frac{16}{3} \zeta_2
          \bigg)

       + C_A C_F^2   \bigg( 
           \frac{4364}{9}
          - 168 \zeta_3
          - \frac{812}{3} \zeta_2
          \bigg)

       + C_A^2 C_F   \bigg( 
          \frac{4676}{27}
          - \frac{98}{3} \zeta_2
          \bigg)\bm{ \bigg\} }

       + {\color{black} \pmb{\overline{\delta} L_1 }}  \bm{ \bigg\{ } C_F n_f^2\bigg( 
          - \frac{1264}{81}
          + \frac{32}{9} \zeta_2
          \bigg)

       + C_F^2 n_f   \bigg( 
           \frac{8290}{81}
          - 32
          - \frac{320}{3} \zeta_3
          - \frac{128}{3} \zeta_2
          \bigg)

       + C_F^3  \bigg( 
          - \frac{262}{3}
          + 848 \zeta_3
          - \frac{352}{3} \zeta_2
          + \frac{192}{5} \zeta_2^2
          \bigg)

       + C_A C_F n_f    \bigg( 
           \frac{18968}{81}
          - \frac{160}{3} 
          - \frac{212}{3} \zeta_2
          \bigg)

       + C_A C_F^2 \   \bigg( 
          - \frac{18562}{81}
          + 568 \zeta_3
          + \frac{788}{3} \zeta_2
          + \frac{48}{5} \zeta_2^2
          \bigg)

       + C_A^2 C_F   \bigg( 
          - \frac{46738}{81}
          + 200 \zeta_3
          + \frac{2560}{9} \zeta_2
          - \frac{176}{5} \zeta_2^2
          \bigg)\bm{ \bigg\} }

       + {\color{black} \pmb{\overline{\delta} L_0 }}   \bm{ \bigg\{ } C_F n_f^2\bigg( 
           \frac{10856}{729}
          + \frac{32}{27} \zeta_3
          - \frac{304}{27} \zeta_2
          \bigg)

       + C_F^2 n_f  \bigg( 
          - \frac{2312}{27}
          + \frac{2032}{9} \zeta_3
          - \frac{232}{9} \zeta_2
          + \frac{256}{15} \zeta_2^2
          \bigg)

       + C_F^3    \bigg( 
           55
          - 384 \zeta_5
          + 112 \zeta_3
          + \frac{256}{3} \zeta_2
          + 128 \zeta_2 \zeta_3
          - \frac{560}{3} \zeta_2^2
          \bigg)

       + C_A C_F n_f   \bigg( 
          - \frac{118984}{729}
          + \frac{196}{3} \zeta_3
          + \frac{11816}{81} \zeta_2
          - \frac{208}{15} \zeta_2^2
          \bigg)

       + C_A C_F^2    \bigg( 
          - \frac{7283}{27}
          - 524 \zeta_3
          - \frac{476}{9} \zeta_2
          + 16 \zeta_2 \zeta_3
          + 24 \zeta_2^2
          \bigg)

       + C_A^2 C_F   \bigg( 
         \frac{576020}{729}
          + 192 \zeta_5
          - \frac{21692}{27} \zeta_3
          - \frac{40844}{81} \zeta_2
          + \frac{176}{3} \zeta_2 \zeta_3
          + \frac{656}{15} \zeta_2^2
          \bigg)\bm{ \bigg\} } 
       
       + {\color{black} \pmb{\overline{{\cal D}}_1 L_2  }}  \bm{ \bigg\{ } C_F^2 n_f\bigg( 
          - \frac{160}{3}
          \bigg)

       + C_F^3  \bigg( 
           416
          \bigg)

       + C_A C_F^2   \bigg( 
           \frac{880}{3}
          \bigg)\bm{ \bigg\} }

       + {\color{black} \pmb{\overline{{\cal D}}_2  L_1}}  \bm{ \bigg\{ } C_F^2 n_f \bigg( 
          - \frac{160}{3}
          \bigg)

       + C_A C_F^2   \bigg( 
           \frac{880}{3}
          \bigg)\bm{ \bigg\} }

       +  {\color{black} \pmb{\overline{{\cal D}}_1  L_1}}  \bm{ \bigg\{ } C_F^2 n_f\bigg( 
           \frac{640}{3}
          \bigg)

       + C_A C_F^2   \bigg( 
          - \frac{3784}{3}
          + 192 \zeta_2
          \bigg)

       + C_A^2 C_F   \bigg( 
          - \frac{968}{9}
          \bigg)\bm{ \bigg\} } 
      
       + {\color{black} \pmb{\overline{{\cal D}}_3  L_0}}  \bm{ \bigg\{ } C_F^2 n_f\bigg( 
          - \frac{160}{9}
          \bigg)

       + C_A C_F^2  \bigg( 
           \frac{880}{9}
          \bigg)\bm{ \bigg\} }

       + {\color{black} \pmb{\overline{{\cal D}}_2 L_0 }}  \bm{ \bigg\{ }  C_F^2 n_f \bigg( 
           96
          \bigg)

       + C_F^3    \bigg( 
           168
          + 96 \zeta_2
          \bigg)

       + C_A^2 C_F   \bigg( 
          - \frac{484}{9}
          \bigg)\bm{ \bigg\} }

       + {\color{black} \pmb{\overline{{\cal D}}_1  L_0}}  \bm{ \bigg\{ }  C_F^2 n_f \bigg( 
          - \frac{296}{3}
          + 64 \zeta_2
          \bigg)

       + C_A C_F n_f   \bigg( 
          - \frac{3896}{27}
          + \frac{32}{3} \zeta_2
          \bigg)

       + C_A^2 C_F  \bigg( 
           \frac{11476}{27}
          - \frac{176}{3} \zeta_2
          \bigg)\bm{ \bigg\} } 

       + {\color{black} \pmb{\overline{{\cal D}}_0 L_3 }}   \bm{ \bigg\{ } C_F^2 n_f\bigg( 
          - \frac{160}{9}
          \bigg)

       + C_F^3  \bigg( 
           160
          \bigg)

       + C_A C_F^2    \bigg( 
          \frac{880}{9}
          \bigg)\bm{ \bigg\} }

       + {\color{black} \pmb{\overline{\delta} L_4}}    \bm{ \bigg\{ } C_F^2 n_f\bigg( 
          - \frac{40}{9}
          \bigg)

     + C_F^3    \bigg( 
           44
          \bigg)

       + C_A C_F^2   \bigg( 
          \frac{220}{9}
          \bigg)\bm{ \bigg\} }

       + {\color{black} \pmb{\overline{{\cal D}}_2 L_2}}    \bm{ \bigg\{ } C_F^3 \bigg( 
          - 240
          \bigg)\bm{ \bigg\} } 

       + {\color{black} \pmb{\overline{{\cal D}}_1 L_3  }}  \bm{ \bigg\{ } C_F^3 \bigg( 
          - 160
          \bigg)\bm{ \bigg\} } 

       + {\color{black} \pmb{\overline{{\cal D}}_3 L_1 }}   \bm{ \bigg\{ } C_F^3 \bigg( 
          - 160
          \bigg)\bm{ \bigg\} } 

       +  {\color{black} \pmb{\overline{{\cal D}}_2  L_1 }}  \bm{ \bigg\{ } C_F^3 \bigg( 
           336
          \bigg)\bm{ \bigg\} }

       + {\color{black} \pmb{\overline{{\cal D}}_4  L_0 }}  \bm{ \bigg\{ } C_F^3 \bigg( 
          - 40
          \bigg)\bm{ \bigg\} } 

       + {\color{black} \pmb{\overline{{\cal D}}_3 L_0 }}  \bm{ \bigg\{ } C_F^3 \bigg( 
           80
          \bigg)\bm{ \bigg\} }

       +{\color{black} \pmb{\overline{{\cal D}}_0 L_4 }}   \bm{ \bigg\{ } C_F^3 \bigg( 
          - 40
          \bigg)\bm{ \bigg\} }

       +{\color{black} \pmb{\overline{\delta} L_5 }}  \bm{ \bigg\{ } C_F^3 \bigg( 
          - 8
          \bigg)\bm{ \bigg\} } 
+
          
          (z_1 \leftrightarrow z_2)\,.
          
\end{autobreak}
\end{align}
Here, $\bm{L_k}= \ln^k(\bar{z}_1)$, $\bm{ \overline{{\cal D}}_j}=\Big[ {\ln^j(\bar{z}_2) \over \bar{z}_2}\Big]  _+$ and $\bm{\overline{\delta} }= \delta(\bar{z}_2)$.
%\end{document}
The NSV exponents $\bar{g}_{d,i}^b$ and $h_{d,ij}^b$ for the rapidity resummation as defined in \eq{eq:PsiN}-\eq{hg}, are given up to the third order as,
\small
\begin{align}
\begin{autobreak}
   \overline{g}_{d,1}^b(\omgot) =

       \frac{1}{\beta_0} C_F   \bigg\{
           2 \Lomg
          \bigg\}\,,

\end{autobreak}\\
  \begin{autobreak}

   \overline{g}_{d,2}^b(\omgot) =

       \frac{1}{1- \omgot} \Bigg[ \frac{1}{\beta_0^2} C_F C_A n_f   \bigg\{
          - \frac{20}{3} \omgot 
          - \frac{20}{3} \Lomg 
          \bigg\}

       +\frac{1}{\beta_0^2} C_F C_A^2   \bigg\{
          \frac{68}{3} \omgot
          + \frac{68}{3} \Lomg
          \bigg\}

       +\frac{1}{\beta_0^2} C_F^2 n_f   \bigg\{
          - 4 \omgot 
          - 4 \Lomg 
          \bigg\}

       +\frac{1}{\beta_0} C_F n_f  \bigg\{
           \frac{20}{9} \omgot 
          \bigg\}

       +\frac{1}{\beta_0} C_F C_A   \bigg\{
          - \frac{134}{9} \omgot
          + 4 \omgot \zeta_2
          \bigg\}

       + C_F   \bigg\{
          - 2
          % + 2 L_{qr}
          % - 2 L_{fr}
          % + 2 L_{fr} \omgot
      
          \bigg\} \Bigg] \,,

\end{autobreak}\\
  \begin{autobreak}

   \overline{g}_{d,3}^b(\omgot) =

       \frac{1}{(1- \omgot)^2} \Bigg[ \frac{1}{\beta_0^3} C_F C_A^2 n_f^2  \bigg\{
           \frac{100}{9} \omgot^2 
          - \frac{100}{9} \Lomg^2 
          \bigg\}

       +\frac{1}{\beta_0^3} C_F C_A^3 n_f   \bigg\{
          - \frac{680}{9} \omgot^2 
          + \frac{680}{9} \Lomg^2 
          \bigg\}

       +\frac{1}{\beta_0^3} C_F C_A^4   \bigg\{
          \frac{1156}{9} \omgot^2
          - \frac{1156}{9} \Lomg^2
          \bigg\}

       +\frac{1}{\beta_0^3} C_F^2 C_A  n_f^2  \bigg\{
          \frac{40}{3} \omgot^2 
          - \frac{40}{3} \Lomg^2 
          \bigg\}

       +\frac{1}{\beta_0^3} C_F^2 C_A^2 n_f   \bigg\{
          - \frac{136}{3} \omgot^2 
          + \frac{136}{3} \Lomg^2 
          \bigg\}

       +\frac{1}{\beta_0^3} C_F^3 n_f^2   \bigg\{
           4 \omgot^2 
          - 4 \Lomg^2 
          \bigg\}

       +\frac{1}{\beta_0^2} C_F C_A n_f^2  \bigg\{
           \frac{200}{27} \omgot 
          - \frac{31}{6} \omgot^2 
          + \frac{200}{27} \Lomg 
          \bigg\}

       +\frac{1}{\beta_0^2} C_F C_A^2 n_f  \bigg\{
          - \frac{2020}{27} \omgot 
          + \frac{40}{3} \omgot \zeta_2 
          + \frac{1145}{18} \omgot^2 
          - \frac{20}{3} \omgot^2 \zeta_2 
          - \frac{2020}{27} \Lomg 
          + \frac{40}{3} \Lomg \zeta_2 
          \bigg\}

       +\frac{1}{\beta_0^2} C_F C_A^3   \bigg\{
           \frac{4556}{27} \omgot
          - \frac{136}{3} \omgot \zeta_2
          - \frac{2471}{18} \omgot^2
          + \frac{68}{3} \omgot^2 \zeta_2
          + \frac{4556}{27} \Lomg
          - \frac{136}{3} \Lomg \zeta_2
          \bigg\}

       +\frac{1}{\beta_0^2} C_F^2 n_f^2  \bigg\{
           \frac{40}{9} \omgot 
          - \frac{31}{9} \omgot^2 
          + \frac{40}{9} \Lomg 
          \bigg\}

       +\frac{1}{\beta_0^2} C_F^2 C_A n_f  \bigg\{
          - \frac{268}{9} \omgot
          + 8 \omgot \zeta_2 
          + \frac{473}{18} \omgot^2 
          - 4 \omgot^2 \zeta_2 
          - \frac{268}{9} \Lomg 
          + 8 \Lomg \zeta_2 
          \bigg\}

       +\frac{1}{\beta_0^2} C_F^3 n_f  \bigg\{
          - \omgot^2 
          \bigg\}

       +\frac{1}{\beta_0} C_F n_f^2  \bigg\{
           \frac{8}{27} \omgot 
          - \frac{4}{27} \omgot^2 
          \bigg\}

       +\frac{1}{\beta_0} C_F C_A n_f   \bigg\{
           \frac{418}{27} \omgot 
          + \frac{56}{3} \omgot  \zeta_3
          - \frac{80}{9} \omgot \zeta_2 
          - \frac{209}{27} \omgot^2 
          - \frac{28}{3} \omgot^2  \zeta_3
          + \frac{40}{9} \omgot^2 \zeta_2 
          - \frac{20}{3} \Lomg 
         % + \frac{20}{3} \Lomg L_{qr} 
       
          \bigg\}

       +\frac{1}{\beta_0} C_F C_A^2   \bigg\{
          - \frac{245}{3} \omgot
          - \frac{44}{3} \omgot \zeta_3
          + \frac{536}{9} \omgot \zeta_2
          - \frac{88}{5} \omgot \zeta_2^2
          + \frac{245}{6} \omgot^2
          + \frac{22}{3} \omgot^2 \zeta_3
          - \frac{268}{9} \omgot^2 \zeta_2
          + \frac{44}{5} \omgot^2 \zeta_2^2
          + \frac{68}{3} \Lomg
        %  - \frac{68}{3} \Lomg L_{qr}
      
          \bigg\}

       +\frac{1}{\beta_0} C_F^2 n_f  \bigg\{
           \frac{55}{3} \omgot 
          - 16 \omgot  \zeta_3
          - \frac{55}{6} \omgot^2 
          + 8 \omgot^2  \zeta_3
          - 4 \Lomg 
       %   + 4 \Lomg L_{qr} 
   
          \bigg\}

       + C_F n_f  \bigg\{
          \frac{116}{27} 
          % - \frac{32}{9} L_{qr}
          % + \frac{2}{3} L_{qr}^2 
          % + \frac{20}{9} L_{fr} 
          % - \frac{40}{9} L_{fr} \omgot 
          % + \frac{20}{9} L_{fr} \omgot^2 
          % - \frac{2}{3} L_{fr}^2 
          % + \frac{4}{3} L_{fr}^2 \omgot 
          % - \frac{2}{3} L_{fr}^2 \omgot^2 
         
          \bigg\}

       + C_F C_A   \bigg\{
          - \frac{806}{27}
          + 14 \zeta_3
          + 4 \zeta_2
          % + \frac{200}{9} L_{qr}
          % - 4 L_{qr} \zeta_2
          % - \frac{11}{3} L_{qr}^2
          % - \frac{134}{9} L_{fr}
          % + 4 L_{fr} \zeta_2
          % + \frac{268}{9} L_{fr} \omgot
          % - 8 L_{fr} \omgot \zeta_2
          % - \frac{134}{9} L_{fr} \omgot^2
          % + 4 L_{fr} \omgot^2 \zeta_2
          % + \frac{11}{3} L_{fr}^2
          % - \frac{22}{3} L_{fr}^2 \omgot
          % + \frac{11}{3} L_{fr}^2 \omgot^2
        
          \bigg\}\Bigg]\,,
\end{autobreak} \\
\begin{autobreak}
  \overline{g}_{d,4}^b(\omgot) =   
  
  \frac{1}{(1-\omgot)^3}\bigg[\frac{1}{\beta_0^2} C_F   \bigg\{
          - \omgot^2 \beta_3
          + \frac{2}{3} \omgot^3 \beta_3
          \bigg\}

       + \frac{\beta_2}{\beta_0^2}  C_F n_f   \bigg\{
           \frac{20}{9} \omgot^2
          - \frac{20}{27} \omgot^3
          \bigg\}

       + \frac{\beta_2}{\beta_0^2}  C_F C_A   \bigg\{
          - \frac{134}{9} \omgot^2
          + 4 \omgot^2 \zeta_2
          + \frac{134}{27} \omgot^3
          - \frac{4}{3} \omgot^3 \zeta_2
          \bigg\}

       + \frac{1}{\beta_0} n_f \frac{d_F^{abcd} d_F^{abcd}}{N_F}   \bigg\{
           \frac{640}{3} \omgot \zeta_5
          + \frac{128}{3} \omgot \zeta_3
          - 128 \omgot \zeta_2
          - \frac{640}{3} \omgot^2 \zeta_5
          - \frac{128}{3} \omgot^2 \zeta_3
          + 128 \omgot^2 \zeta_2
          + \frac{640}{9} \omgot^3 \zeta_5
          + \frac{128}{9} \omgot^3 \zeta_3
          - \frac{128}{3} \omgot^3 \zeta_2
          \bigg\}

       + \frac{1}{\beta_0} \frac{d_F^{abcd} d_A^{abcd}}{N_F}   \bigg\{
          - \frac{1760}{3} \omgot \zeta_5
          - \frac{64}{3} \omgot \zeta_3
          + 192 \omgot \zeta_3^2
          + 64 \omgot \zeta_2
          + \frac{3968}{35} \omgot \zeta_2^3
          + \frac{1760}{3} \omgot^2 \zeta_5
          + \frac{64}{3} \omgot^2 \zeta_3
          - 192 \omgot^2 \zeta_3^2
          - 64 \omgot^2 \zeta_2
          - \frac{3968}{35} \omgot^2 \zeta_2^3
          - \frac{1760}{9} \omgot^3 \zeta_5
          - \frac{64}{9} \omgot^3 \zeta_3
          + 64 \omgot^3 \zeta_3^2
          + \frac{64}{3} \omgot^3 \zeta_2
          + \frac{3968}{105} \omgot^3 \zeta_2^3
          \bigg\}

       + \frac{1}{\beta_0} C_F n_f^3   \bigg\{
           \frac{16}{81} \omgot
          - \frac{32}{27} \omgot \zeta_3
          - \frac{16}{81} \omgot^2
          + \frac{32}{27} \omgot^2 \zeta_3
          + \frac{16}{243} \omgot^3
          - \frac{32}{81} \omgot^3 \zeta_3
          \bigg\}

       + \frac{1}{\beta_0} C_F C_A n_f^2   \bigg\{
          - \frac{923}{162} \omgot
          - \frac{1120}{27} \omgot \zeta_3
          + \frac{304}{81} \omgot \zeta_2
          + \frac{112}{15} \omgot \zeta_2^2
          + \frac{923}{162} \omgot^2
          + \frac{1120}{27} \omgot^2 \zeta_3
          - \frac{304}{81} \omgot^2 \zeta_2
          - \frac{112}{15} \omgot^2 \zeta_2^2
          - \frac{923}{486} \omgot^3
          - \frac{1120}{81} \omgot^3 \zeta_3
          + \frac{304}{243} \omgot^3 \zeta_2
          + \frac{112}{45} \omgot^3 \zeta_2^2
          \bigg\}

       + \frac{1}{\beta_0} C_F C_A^2 n_f   \bigg\{
           \frac{24137}{162} \omgot
          - \frac{1048}{9} \omgot \zeta_5
          + \frac{11552}{27} \omgot \zeta_3
          - \frac{10160}{81} \omgot \zeta_2
          - \frac{224}{3} \omgot \zeta_2 \zeta_3
          + \frac{176}{15} \omgot \zeta_2^2
          - \frac{24137}{162} \omgot^2
          + \frac{1048}{9} \omgot^2 \zeta_5
          - \frac{11552}{27} \omgot^2 \zeta_3
          + \frac{10160}{81} \omgot^2 \zeta_2
          + \frac{224}{3} \omgot^2 \zeta_2 \zeta_3
          - \frac{176}{15} \omgot^2 \zeta_2^2
          + \frac{24137}{486} \omgot^3
          - \frac{1048}{27} \omgot^3 \zeta_5
          + \frac{11552}{81} \omgot^3 \zeta_3
          - \frac{10160}{243} \omgot^3 \zeta_2
          - \frac{224}{9} \omgot^3 \zeta_2 \zeta_3
          + \frac{176}{45} \omgot^3 \zeta_2^2
          \bigg\}

       + \frac{1}{\beta_0} C_F C_A^3   \bigg\{
          - \frac{42139}{81} \omgot
          + \frac{1804}{9} \omgot \zeta_5
          - \frac{10472}{27} \omgot \zeta_3
          + 8 \omgot \zeta_3^2
          + \frac{44200}{81} \omgot \zeta_2
          + \frac{176}{3} \omgot \zeta_2 \zeta_3
          - \frac{1804}{5} \omgot \zeta_2^2
          + \frac{10016}{105} \omgot \zeta_2^3
          + \frac{42139}{81} \omgot^2
          - \frac{1804}{9} \omgot^2 \zeta_5
          + \frac{10472}{27} \omgot^2 \zeta_3
          - 8 \omgot^2 \zeta_3^2
          - \frac{44200}{81} \omgot^2 \zeta_2
          - \frac{176}{3} \omgot^2 \zeta_2 \zeta_3
          + \frac{1804}{5} \omgot^2 \zeta_2^2
          - \frac{10016}{105} \omgot^2 \zeta_2^3
          - \frac{42139}{243} \omgot^3
          + \frac{1804}{27} \omgot^3 \zeta_5
          - \frac{10472}{81} \omgot^3 \zeta_3
          + \frac{8}{3} \omgot^3 \zeta_3^2
          + \frac{44200}{243} \omgot^3 \zeta_2
          + \frac{176}{9} \omgot^3 \zeta_2 \zeta_3
          - \frac{1804}{15} \omgot^3 \zeta_2^2
          + \frac{10016}{315} \omgot^3 \zeta_2^3
          \bigg\}

       + \frac{1}{\beta_0} C_F^2 n_f^2   \bigg\{
          - \frac{1196}{81} \omgot
          + \frac{320}{9} \omgot \zeta_3
          - \frac{32}{5} \omgot \zeta_2^2
          + \frac{1196}{81} \omgot^2
          - \frac{320}{9} \omgot^2 \zeta_3
          + \frac{32}{5} \omgot^2 \zeta_2^2
          - \frac{1196}{243} \omgot^3
          + \frac{320}{27} \omgot^3 \zeta_3
          - \frac{32}{15} \omgot^3 \zeta_2^2
          \bigg\}

       + \frac{1}{\beta_0} C_F^2 C_A n_f   \bigg\{
           \frac{17033}{81} \omgot
          - 80 \omgot \zeta_5
          - \frac{1856}{9} \omgot \zeta_3
          - \frac{220}{3} \omgot \zeta_2
          + 64 \omgot \zeta_2 \zeta_3
          + \frac{176}{5} \omgot \zeta_2^2
          - \frac{17033}{81} \omgot^2
          + 80 \omgot^2 \zeta_5
          + \frac{1856}{9} \omgot^2 \zeta_3
          + \frac{220}{3} \omgot^2 \zeta_2
          - 64 \omgot^2 \zeta_2 \zeta_3
          - \frac{176}{5} \omgot^2 \zeta_2^2
          + \frac{17033}{243} \omgot^3
          - \frac{80}{3} \omgot^3 \zeta_5
          - \frac{1856}{27} \omgot^3 \zeta_3
          - \frac{220}{9} \omgot^3 \zeta_2
          + \frac{64}{3} \omgot^3 \zeta_2 \zeta_3
          + \frac{176}{15} \omgot^3 \zeta_2^2
          \bigg\}

       + \frac{1}{\beta_0} C_F^3 n_f   \bigg\{
          - \frac{286}{9} \omgot
          + 160 \omgot \zeta_5
          - \frac{296}{3} \omgot \zeta_3
          + \frac{286}{9} \omgot^2
          - 160 \omgot^2 \zeta_5
          + \frac{296}{3} \omgot^2 \zeta_3
          - \frac{286}{27} \omgot^3
          + \frac{160}{3} \omgot^3 \zeta_5
          - \frac{296}{9} \omgot^3 \zeta_3
          \bigg\}

       + \frac{\beta_2}{\beta_0}  C_F   \bigg\{
          - 2 \omgot
         % + 2 L_{qr} \omgot
          \bigg\}

       + C_F n_f^2   \bigg\{
           \frac{1256}{729}
          - \frac{56}{27} \zeta_3
          + \frac{20}{27} \zeta_2
          % - \frac{8}{27} L_{qr}
          % + \frac{8}{27} L_{fr}
          % - \frac{8}{9} L_{fr} \omgot
          % + \frac{8}{9} L_{fr} \omgot^2
          % - \frac{8}{27} L_{fr} \omgot^3
          \bigg\}

       + C_F C_A n_f   \bigg\{
           \frac{17207}{729}
          + \frac{140}{27} \zeta_3
          - \frac{2134}{81} \zeta_2
          + \frac{48}{5} \zeta_2^2
          % - \frac{418}{27} L_{qr}
          % - \frac{56}{3} L_{qr} \zeta_3
          % + \frac{80}{9} L_{qr} \zeta_2
          % + \frac{418}{27} L_{fr}
          % + \frac{56}{3} L_{fr} \zeta_3
          % - \frac{80}{9} L_{fr} \zeta_2
          % - \frac{418}{9} L_{fr} \omgot
          % - 56 L_{fr} \omgot \zeta_3
          % + \frac{80}{3} L_{fr} \omgot \zeta_2
          % + \frac{418}{9} L_{fr} \omgot^2
          % + 56 L_{fr} \omgot^2 \zeta_3
          % - \frac{80}{3} L_{fr} \omgot^2 \zeta_2
          % - \frac{418}{27} L_{fr} \omgot^3
          % - \frac{56}{3} L_{fr} \omgot^3 \zeta_3
          % + \frac{80}{9} L_{fr} \omgot^3 \zeta_2
          \bigg\}

       + C_F C_A^2   \bigg\{
          - \frac{255851}{1458}
          - 96 \zeta_5
          + \frac{614}{3} \zeta_3
          + \frac{11149}{81} \zeta_2
          - \frac{88}{3} \zeta_2 \zeta_3
          - \frac{264}{5} \zeta_2^2
          % + \frac{245}{3} L_{qr}
          % + \frac{44}{3} L_{qr} \zeta_3
          % - \frac{536}{9} L_{qr} \zeta_2
          % + \frac{88}{5} L_{qr} \zeta_2^2
          % - \frac{245}{3} L_{fr}
          % - \frac{44}{3} L_{fr} \zeta_3
          % + \frac{536}{9} L_{fr} \zeta_2
          % - \frac{88}{5} L_{fr} \zeta_2^2
          % + 245 L_{fr} \omgot
          % + 44 L_{fr} \omgot \zeta_3
          % - \frac{536}{3} L_{fr} \omgot \zeta_2
          % + \frac{264}{5} L_{fr} \omgot \zeta_2^2
          % - 245 L_{fr} \omgot^2
          % - 44 L_{fr} \omgot^2 \zeta_3
          % + \frac{536}{3} L_{fr} \omgot^2 \zeta_2
          % - \frac{264}{5} L_{fr} \omgot^2 \zeta_2^2
          % + \frac{245}{3} L_{fr} \omgot^3
          % + \frac{44}{3} L_{fr} \omgot^3 \zeta_3
          % - \frac{536}{9} L_{fr} \omgot^3 \zeta_2
          % + \frac{88}{5} L_{fr} \omgot^3 \zeta_2^2
          \bigg\}

       + C_F^2 n_f   \bigg\{
           \frac{2701}{54}
          - \frac{296}{9} \zeta_3
          - 2 \zeta_2
          - \frac{16}{5} \zeta_2^2
          % - \frac{55}{3} L_{qr}
          % + 16 L_{qr} \zeta_3
          % + \frac{55}{3} L_{fr}
          % - 16 L_{fr} \zeta_3
          % - 55 L_{fr} \omgot
          % + 48 L_{fr} \omgot \zeta_3
          % + 55 L_{fr} \omgot^2
          % - 48 L_{fr} \omgot^2 \zeta_3
          % - \frac{55}{3} L_{fr} \omgot^3
          % + 16 L_{fr} \omgot^3 \zeta_3
          \bigg\}

       + \beta_0 C_F n_f   \bigg\{
           \frac{664}{81}
          - \frac{32}{3} \zeta_3
          - \frac{2}{9} \zeta_2
          % - \frac{232}{27} L_{qr}
          % + \frac{4}{3} L_{qr} \zeta_2
          % + \frac{20}{9} L_{qr}^2
          % - \frac{20}{9} L_{fr}^2
          % + \frac{20}{3} L_{fr}^2 \omgot
          % - \frac{20}{3} L_{fr}^2 \omgot^2
          % + \frac{20}{9} L_{fr}^2 \omgot^3
          \bigg\}

       + \beta_0 C_F C_A   \bigg\{
          - \frac{4852}{81}
          + \frac{260}{3} \zeta_3
          - \frac{1}{9} \zeta_2
          + 12 \zeta_2^2
          % + \frac{1612}{27} L_{qr}
          % - 28 L_{qr} \zeta_3
          % - \frac{46}{3} L_{qr} \zeta_2
          % - \frac{134}{9} L_{qr}^2
          % + 4 L_{qr}^2 \zeta_2
          % + \frac{134}{9} L_{fr}^2
          % - 4 L_{fr}^2 \zeta_2
          % - \frac{134}{3} L_{fr}^2 \omgot
          % + 12 L_{fr}^2 \omgot \zeta_2
          % + \frac{134}{3} L_{fr}^2 \omgot^2
          % - 12 L_{fr}^2 \omgot^2 \zeta_2
          % - \frac{134}{9} L_{fr}^2 \omgot^3
          % + 4 L_{fr}^2 \omgot^3 \zeta_2
          \bigg\}

       + \beta_0^2 C_F   \bigg\{
          - \frac{46}{3} \zeta_3
          - 2 \zeta_2
          % + 2 L_{qr} \zeta_2
          % - 2 L_{qr}^2
          % + \frac{2}{3} L_{qr}^3
          % - \frac{2}{3} L_{fr}^3
          % + 2 L_{fr}^3 \omgot
          % - 2 L_{fr}^3 \omgot^2
          % + \frac{2}{3} L_{fr}^3 \omgot^3
          \bigg\}

       + \frac{\beta_1}{\beta_0^3} \beta_2 C_F   \bigg\{
         2 \omgot^2
          - \frac{4}{3} \omgot^3
          + 2 \Lomg \omgot
          \bigg\}

       + \frac{\beta_1}{\beta_0^2} C_F n_f^2   \bigg\{
          - \frac{8}{27} \omgot
          + \frac{8}{27} \omgot^2
          - \frac{8}{81} \omgot^3
          - \frac{8}{27} \Lomg
          \bigg\}

       + \frac{\beta_1}{\beta_0^2} C_F C_A n_f   \bigg\{
          - \frac{418}{27} \omgot
          - \frac{56}{3} \omgot \zeta_3
          + \frac{80}{9} \omgot \zeta_2
          + \frac{418}{27} \omgot^2
          + \frac{56}{3} \omgot^2 \zeta_3
          - \frac{80}{9} \omgot^2 \zeta_2
          - \frac{418}{81} \omgot^3
          - \frac{56}{9} \omgot^3 \zeta_3
          + \frac{80}{27} \omgot^3 \zeta_2
          - \frac{418}{27} \Lomg
          - \frac{56}{3} \Lomg \zeta_3
          + \frac{80}{9} \Lomg \zeta_2
          \bigg\}

       +  \frac{\beta_1}{\beta_0^2} C_F C_A^2   \bigg\{
           \frac{245}{3} \omgot
          + \frac{44}{3} \omgot \zeta_3
          - \frac{536}{9} \omgot \zeta_2
          + \frac{88}{5} \omgot \zeta_2^2
          - \frac{245}{3} \omgot^2
          - \frac{44}{3} \omgot^2 \zeta_3
          + \frac{536}{9} \omgot^2 \zeta_2
          - \frac{88}{5} \omgot^2 \zeta_2^2
          + \frac{245}{9} \omgot^3
          + \frac{44}{9} \omgot^3 \zeta_3
          - \frac{536}{27} \omgot^3 \zeta_2
          + \frac{88}{15} \omgot^3 \zeta_2^2
          + \frac{245}{3} \Lomg
          + \frac{44}{3} \Lomg \zeta_3
          - \frac{536}{9} \Lomg \zeta_2
          + \frac{88}{5} \Lomg \zeta_2^2
          \bigg\}

       +  \frac{\beta_1}{\beta_0^2} C_F^2 n_f   \bigg\{
          - \frac{55}{3} \omgot
          + 16 \omgot \zeta_3
          + \frac{55}{3} \omgot^2
          - 16 \omgot^2 \zeta_3
          - \frac{55}{9} \omgot^3
          + \frac{16}{3} \omgot^3 \zeta_3
          - \frac{55}{3} \Lomg
          + 16 \Lomg \zeta_3
          \bigg\}

       +  \frac{\beta_1}{\beta_0} C_F n_f   \bigg\{
          - \frac{232}{27} \Lomg
          + \frac{4}{3} \Lomg \zeta_2
        %  + \frac{40}{9} \Lomg L_{qr}
          \bigg\}

       + \frac{\beta_1}{\beta_0} C_F C_A   \bigg\{
           \frac{1612}{27} \Lomg
          - 28 \Lomg \zeta_3
          - \frac{46}{3} \Lomg \zeta_2
          % - \frac{268}{9} \Lomg L_{qr}
          % + 8 \Lomg L_{qr} \zeta_2
          \bigg\}

       + \beta_1 C_F   \bigg\{
          - \zeta_2
          % + 2 L_{qr}
          % - L_{qr}^2
          + 2 \Lomg \zeta_2
          % - 4 \Lomg L_{qr}
          % + 2 \Lomg L_{qr}^2
          % + L_{fr}^2
          % - 3 L_{fr}^2 \omgot
          % + 3 L_{fr}^2 \omgot^2
          % - L_{fr}^2 \omgot^3
          \bigg\}

       +  \frac{\beta_1^2}{\beta_0^3} C_F n_f   \bigg\{
          - \frac{20}{9} \omgot^2
          + \frac{20}{27} \omgot^3
          + \frac{20}{9} \Lomg^2
          \bigg\}

       +  \frac{\beta_1^2}{\beta_0^3} C_F C_A   \bigg\{
           \frac{134}{9} \omgot^2
          - 4 \omgot^2 \zeta_2
          - \frac{134}{27} \omgot^3
          + \frac{4}{3} \omgot^3 \zeta_2
          - \frac{134}{9} \Lomg^2
          + 4 \Lomg^2 \zeta_2
          \bigg\}

       + \frac{\beta_1^2}{\beta_0^2} C_F   \bigg\{
           2 \omgot
        %  - 2 L_{qr} \omgot
          + 2 \Lomg
        %  - 2 \Lomg L_{qr}
          - 2 \Lomg^2
        %  + 2 \Lomg^2 L_{qr}
          \bigg\}

       +  \frac{\beta_1^3}{\beta_0^4} C_F   \bigg\{
          - \omgot^2
          + \frac{2}{3} \omgot^3
          - 2 \Lomg \omgot
          - \Lomg^2
          + \frac{2}{3} \Lomg^3
          \bigg\} \bigg]\,,

\end{autobreak}
\end{align}
%------------------------------------------------------------
\small

Here, $L_{\omega} = \ln(1-\omgot)$ with $\omgot = a_s(\mu_R^2) \beta_0 \ln(\overbar{N}_1 \overline{N}_2)$ and  $\omega_l = a_s(\mu_R^2) \beta_0 \ln(\overbar{N}_l)$.

%%%References
\bibliographystyle{JHEP}
\bibliography{bbHnsv} 
\end{document}